\documentclass[sigconf,screen]{acmart}

\usepackage[utf8]{inputenc}
\usepackage[T1]{fontenc}
\usepackage{hyphenat}
\usepackage{xspace}
\usepackage{amsmath}
\usepackage{amsfonts}
\usepackage{hyperref}
\usepackage{url}
\usepackage{booktabs}
\usepackage{multirow}
\usepackage{makecell}
\usepackage{caption}
\usepackage{minibox}
\usepackage{bbm}
\usepackage{graphicx}
\usepackage{balance}
\usepackage{mathtools}
\usepackage{color}
\usepackage{marvosym}
\usepackage{ifthen}
\usepackage{textcomp}
\usepackage{enumitem}
\usepackage{verbatim}
\usepackage{algorithm}
\usepackage{algorithmic}
\usepackage{numprint}
\usepackage{balance}

\usepackage[export]{adjustbox}
\usepackage{subcaption}

\usepackage{amsthm}
\theoremstyle{plain}

\newcommand{\chatoDisplayMode}[1]{#1}

\definecolor{MyRed}{rgb}{0.6,0.0,0.0} 
\definecolor{MyBlack}{rgb}{0.1,0.1,0.1} 
\newcommand{\inred}[1]{{\color{MyRed}\sf\textbf{\textsc{#1}}}}
\newcommand{\frameit}[2]{
  \begin{center}
  {\color{MyRed}
  \framebox[.9\columnwidth][l]{
    \begin{minipage}{.85\columnwidth}
    \inred{#1}: {\sf\color{MyBlack}#2}
    \end{minipage}
  }\\
  }
  \end{center}
}

\newcommand{\note}[2][]{\chatoDisplayMode{\def\@tmpsig{#1}\frameit{{\Pointinghand} Note}{#2\ifx \@tmpsig \@empty \else \mbox{ --\em #1}\fi}}}
\newcommand{\todo}[2][]{\chatoDisplayMode{\def\@tmpsig{#1}\frameit{{\Writinghand} To-do}{#2\ifx \@tmpsig \@empty \else \mbox{ --\em #1}\fi}}}

\newcommand{\abbrevStyle}[1]{#1}

\newcommand{\ie}{\abbrevStyle{i.e.}\xspace}
\newcommand{\eg}{\abbrevStyle{e.g.}\xspace}
\newcommand{\cf}{\abbrevStyle{cf.}\xspace}

\newcommand{\vs}{\abbrevStyle{vs.}\xspace}
\newcommand{\etc}{\abbrevStyle{etc.}\xspace}

\newcommand{\Secref}[1]{Sec.~\ref{#1}}

\newcommand{\Tabref}[1]{Table~\ref{#1}}
\newcommand{\Figref}[1]{Fig.~\ref{#1}}

\newcommand{\xhdr}[1]{\vspace{1.7mm}\noindent{{\bf #1.}}}

\newcommand{\textcite}[1]{\citeauthor{#1} \shortcite{#1}}

\newcommand{\cpt}[1]{\textsc{\MakeLowercase{#1}}}

\newcommand{\hide}[1]{}

\makeatletter
\newcommand{\iffont}[2]{\ifthenelse{\equal{\f@family}{#1}}{#2}{}}
\makeatother

\iffont{ptm}{
  \usepackage{mathptmx}

  \DeclareSymbolFont{greek}{OML}{cmm}{m}{n}
  \DeclareMathSymbol{\alpha}{\mathalpha}{greek}{"0B}
  \DeclareMathSymbol{\beta}{\mathalpha}{greek}{"0C}
  \DeclareMathSymbol{\gamma}{\mathalpha}{greek}{"0D}
  \DeclareMathSymbol{\delta}{\mathalpha}{greek}{"0E}
  \DeclareMathSymbol{\epsilon}{\mathalpha}{greek}{"0F}
  \DeclareMathSymbol{\zeta}{\mathalpha}{greek}{"10}
  \DeclareMathSymbol{\eta}{\mathalpha}{greek}{"11}
  \DeclareMathSymbol{\theta}{\mathalpha}{greek}{"12}
  \DeclareMathSymbol{\iota}{\mathalpha}{greek}{"13}
  \DeclareMathSymbol{\kappa}{\mathalpha}{greek}{"14}
  \DeclareMathSymbol{\lambda}{\mathalpha}{greek}{"15}
  \DeclareMathSymbol{\mu}{\mathalpha}{greek}{"16}
  \DeclareMathSymbol{\nu}{\mathalpha}{greek}{"17}
  \DeclareMathSymbol{\xi}{\mathalpha}{greek}{"18}
  \DeclareMathSymbol{\pi}{\mathalpha}{greek}{"19}
  \DeclareMathSymbol{\rho}{\mathalpha}{greek}{"1A}
  \DeclareMathSymbol{\sigma}{\mathalpha}{greek}{"1B}
  \DeclareMathSymbol{\tau}{\mathalpha}{greek}{"1C}
  \DeclareMathSymbol{\upsilon}{\mathalpha}{greek}{"1D}
  \DeclareMathSymbol{\phi}{\mathalpha}{greek}{"1E}
  \DeclareMathSymbol{\chi}{\mathalpha}{greek}{"1F}
  \DeclareMathSymbol{\psi}{\mathalpha}{greek}{"20}
  \DeclareMathSymbol{\omega}{\mathalpha}{greek}{"21}
  \DeclareMathSymbol{\varepsilon}{\mathalpha}{greek}{"22}
  \DeclareMathSymbol{\vartheta}{\mathalpha}{greek}{"23}
  \DeclareMathSymbol{\varpi}{\mathalpha}{greek}{"24}
  \DeclareMathSymbol{\varrho}{\mathalpha}{greek}{"25}
  \DeclareMathSymbol{\varsigma}{\mathalpha}{greek}{"26}
  \DeclareMathSymbol{\varphi}{\mathalpha}{greek}{"27}
  \DeclareSymbolFont{otone}{OT1}{cmr}{m}{n}
  \DeclareMathSymbol{\Gamma}{\mathalpha}{otone}{0}
  \DeclareMathSymbol{\Delta}{\mathalpha}{otone}{1}
  \DeclareMathSymbol{\Theta}{\mathalpha}{otone}{2}
  \DeclareMathSymbol{\Lambda}{\mathalpha}{otone}{3}
  \DeclareMathSymbol{\Xi}{\mathalpha}{otone}{4}
  \DeclareMathSymbol{\Pi}{\mathalpha}{otone}{5}
  \DeclareMathSymbol{\Sigma}{\mathalpha}{otone}{6}
  \DeclareMathSymbol{\Upsilon}{\mathalpha}{otone}{7}
  \DeclareMathSymbol{\Phi}{\mathalpha}{otone}{8}
  \DeclareMathSymbol{\Psi}{\mathalpha}{otone}{9}
  \DeclareMathSymbol{\Omega}{\mathalpha}{otone}{10}
  \DeclareSymbolFont{syms}{OML}{cmm}{m}{it}
  \DeclareMathSymbol{\partial}{\mathord}{syms}{"40}
  \DeclareMathAlphabet{\mathbold}{OML}{cmm}{b}{it}
  \DeclareSymbolFont{largesymbols}{OMX}{cmex}{m}{n}

}

\usepackage{makecell}
\usepackage{wrapfig}

\newcommand{\WP}{Wikipedia\xspace}
\newcommand{\ctr}{click-through rate\xspace}
\newcommand{\Ctr}{Click-through rate\xspace}

\definecolor{airforceblue}{rgb}{0.2, 0.2, 0.66}

\copyrightyear{2021}
\acmYear{2021}
\setcopyright{iw3c2w3}
\acmConference[WWW '21]{Proceedings of The Web Conference 2021}{April 19--23, 2021}{Ljubljana, Slovenia}
\acmBooktitle{Proceedings of The Web Conference 2021 (WWW '21), April 19--23, 2021, Ljubljana, Slovenia}
\acmPrice{}
\acmDOI{10.1145/3442381.3450136}
\acmISBN{978-1-4503-8312-7/21/04}

\begin{document}


\title{On the Value of Wikipedia as a Gateway to the Web}



\author{Tiziano Piccardi}
\affiliation{%
  \institution{EPFL}
}
\email{tiziano.piccardi@epfl.ch}

\author{Miriam Redi}
\affiliation{%
  \institution{Wikimedia Foundation}
}
\email{miriam@wikimedia.org}

\author{Giovanni Colavizza}
\affiliation{%
  \institution{University of Amsterdam}
}
\email{g.colavizza@uva.nl}

\author{Robert West}
\authornote{Robert West is a Wikimedia Foundation Research Fellow.}
\affiliation{%
  \institution{EPFL}
}
\email{robert.west@epfl.ch}



\begin{abstract}
By linking to external websites, Wikipedia can act as a gateway to the Web. To date, however, little is known about the amount of traffic generated by Wikipedia's external links. We fill this gap in a detailed analysis of usage logs gathered from Wikipedia users' client devices. Our analysis proceeds in three steps:
First, we quantify the level of engagement with external links, finding that, in one month, English Wikipedia generated 43M clicks to external websites, in roughly even parts via links in infoboxes, cited references, and article bodies. Official links listed in infoboxes have by far the highest click-through rate (CTR), 2.47\% on average. In particular, official links associated with articles about businesses, educational institutions, and websites have the highest CTR, whereas official links associated with articles about geographical content, television, and music have the lowest CTR.
Second, we investigate patterns of engagement with external links, finding that Wikipedia frequently serves as a stepping stone between search engines and third-party websites, effectively fulfilling information needs that search engines do not meet.
Third, we quantify the hypothetical economic value of the clicks received by external websites from English Wikipedia, by estimating that the respective website owners would need to pay a total of \$7--13 million per month to obtain the same volume of traffic via sponsored search.
Overall, these findings shed light on Wikipedia's role not only as an important source of information, but also as a high-traffic gateway to the broader Web ecosystem. 
\end{abstract}

\maketitle



\section{Introduction}

Thanks to the collaborative effort of a community of volunteer editors, Wikipedia is the world's largest encyclopedia and an important source of information for millions of people.
Wikipedia serves its content as a regular website, allowing editors to add hyperlinks in
\begin{wrapfigure}{r}{0.5\columnwidth}
  \begin{center}
  \vspace{-1.5mm}
  \includegraphics[width=0.5\columnwidth]{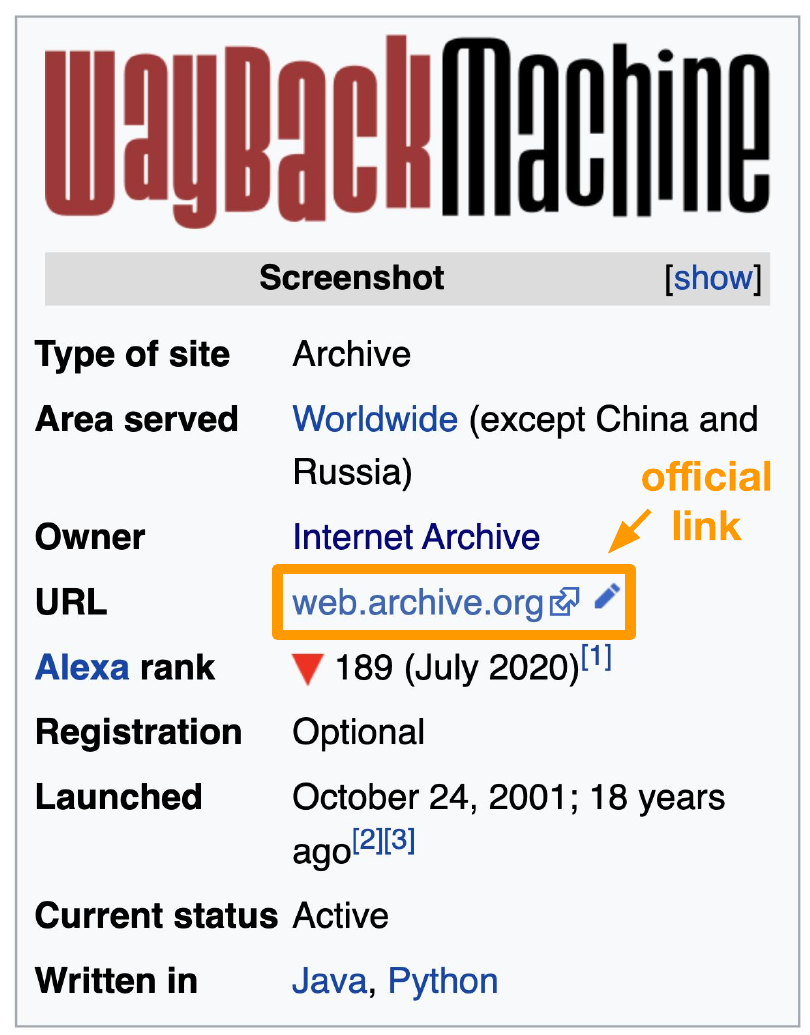}
  \vspace{-6mm}
  \end{center}
  \caption{Example of an official link, in infobox of Wikipedia article about Internet Archive's Wayback Machine.}
  \vspace{-1mm}
  \label{fig:screenshot}
\end{wrapfigure}
order to enable readers to more easily find additional content, both internal and external.
Internal links help readers locate relevant encyclopedic content by navigating from article to article.
In contrast, external links enrich articles with additional content that should not or cannot be included in Wikipedia itself.
There are various reasons to add external links,%
\footnote{\url{https://en.wikipedia.org/wiki/Wikipedia:External_links}}
with linked content ranging from official websites, to news articles used as references,%
\footnote{\url{https://en.wikipedia.org/wiki/Wikipedia:Citing_sources}}
to copyrighted material.

In this paper, we are interested in quantifying and characterizing the outgoing traffic generated by Wikipedia towards external content.
Given \WP's crucial societal role and global reach, it is essential to understand how it interacts with the broader Web by driving traffic to external websites.
The resulting insights can inform the platform's future design and thus allow it to better cater to readers' information needs around external content.
As Web traffic has monetary value---in particular when the traffic goes to commercial websites---an investigation of the external traffic generated by \WP also sheds new light on the poorly understood role it has as a provider not only of information, but also of economic wealth.

\xhdr{Research questions}
We approach the question of \WP's value as a gateway to the Web from two angles: informational and economic.
Concretely, we pose three research questions:

\begin{description}
    \item[RQ1] \textbf{Level of engagement with external links:} What total volume of traffic does \WP drive to third-party websites? What is the \ctr of external links, and how does it vary across types of linked content? (\Secref{sec:RQ1})
    \item[RQ2] \textbf{Patterns of engagement with external links:} How do users interact with external links? Do they click through fast or slow, and how does this vary across types of linked content? In what navigational situations do clicks to external websites occur? (\Secref{sec:RQ2})
    \item[RQ3] \textbf{Economic value of external links:} What is the monetary value of the traffic from Wikipedia to external websites? If website owners had to pay for an equivalent amount of traffic via sponsored search, how much would this cost? (\Secref{sec:RQ3})
\end{description}

\xhdr{Summary of findings}
Based on usage logs gathered over a one-month period from English \WP users' client devices (\Secref{sec:Data}),
we quantified the level of engagement with external links \textbf{(RQ1)}, determining that English \WP generated 43 million clicks to external websites during the month we studied, despite the fact that, on average, the \ctr (CTR) of external links was only 0.08\%.
While most external links (95.5\%) occurred in article bodies and cited references (accounting for about two thirds of the external traffic),
a disproportionately large fraction (23\%) of the total traffic came from a relatively small fraction (0.8\%) of all external links, namely from \textit{official links} to the website of the entity covered in the respective article.
Such official links are regularly listed in so-called \textit{infoboxes,} short tabular summaries of key facts about the covered entity (see \Figref{fig:screenshot} for an example).
Since official links witnessed a vastly increased CTR of 2.47\% (\vs\ 0.08\% over all external links), we focused our analysis on official links.
In a topical analysis, we found that official links associated with articles about businesses, educational institutions, and websites had the highest CTR (a first indicator of the economic value of \WP's external links), whereas official links associated with articles about geographical content, television, and music had the lowest CTR.

By analyzing patterns of engagement with external links \textbf{(RQ2),} we observed that Wikipedia frequently serves as a stepping stone between search engines and third-party websites.
We captured this effect quantitatively as well as in a manual analysis, where we found that URLs that are down\hyp ranked or censored by search engines, and thus not retrievable via search, can often be found in \WP infoboxes, which leads search users to take a detour via \WP.
We conclude that \WP regularly and systematically meets information needs that search engines do not meet, which further confirms \WP's central role in the Web ecosystem.

Finally, we aimed to quantify the hypothetical economic value of the clicks received by external websites from English Wikipedia \textbf{(RQ3).}
Wikipedia is, of course, free, and it runs thanks to the donations of thousands of people.
We thus cannot ask how much money \WP could earn by charging a fee for external clicks---this hypothetical scenario is simply too far from reality---but we may approach the question from a different angle, asking how much money external\hyp website owners would have to pay in order to obtain an equivalent number of clicks by other means, such as paid ads.
In this spirit, we applied the Google Ads API to the content of official websites linked from \WP in order to generate keywords for sponsored search and estimated their cost per click at market price.
We conclude that the owners of external websites linked from English \WP's infoboxes would need to collectively pay a total of around \$7--13 million per month (or \$84--156 million per year) for sponsored search in order to obtain the same volume of traffic that they receive from \WP for free.

These numbers exceed even the ballpark guess given in a bullish 2013 analysis~\cite{johnston_2013} that, unlike ours, was not based on real click logs, but on generic rates commonly assumed in the online ad industry, and estimated that \WP could earn \$2.5 million monthly via affiliate links.
Although our analysis of monetary value should mostly be taken as an indicative ``back-of-the\hyp envelope'' calculation,
it highlights the importance of \WP not only as a source of information, but also as a gratuitous provider of economic wealth.


\begin{table*}[t]
  \centering
  \caption{
  Click statistics for external links embedded in Wikipedia articles.
  }
    \vspace{-2mm}
\begin{tabular}{l|rr|rr|rrrr}
\textbf{Link location} & \multicolumn{2}{c|}{\textbf{Total links}} & \multicolumn{2}{c|}{\textbf{Clicks}} & \textbf{Total} & \textbf{Mean} & \textbf{Median} \\

 & Number & Perc. of total & Number & Perc. of total  & \textbf{articles}  &  \textbf{CTR $\pm$ SD*} & \textbf{click time\textsuperscript{\textdagger}} \\
 \hline

Infobox  & 2.8M & 4.5\% & 12.5M & 29.1\% & 1.3M &  0.90\% $\pm$ 2.2\% & 18.6s (45.6s)\\
\hspace{3mm}\textit{Official links} & \textit{506K} & \textit{0.8\%} & \textit{9.8M} & \textit{22.7\%} & \textit{506K}  &\textit{ 2.47\% $\pm$ 3.0\%} & \textit{20.7s (47.8s)}\\
Body & 24.9M & 39.5\% & 16.2M & 37.8\% & 4.0M & 0.14\% $\pm$ 0.7\% & 35.4s (90.9s)\\
References & 35.3M & 56.0\% & 14.2M & 33.1\% & 3.9M &  0.03\% $\pm$ 0.2\% & 51.8s (131.4s)\\
\hline

 All links & 63.1M & 100\% & 43.1M & 100\% &  &  0.08\% $\pm$ 0.5\% & 32.9s (87.1s)
 \\
 \hline
\end{tabular}
\center
{\small
  *CTR = \ctr; considering only links with at least 300 impressions during the one-month study period.
  \textsuperscript{\textdagger}Inter-quartile range in parentheses.
}
  \label{tab:general_stats}
\end{table*}

\section{Related work}

\xhdr{User engagement on the Web} Being able to quantify user engagement is crucial for websites, especially for those with an advertising\hyp based business model~\cite{user_engagement}.
Researchers from various fields have investigated ways to measure users' attention, interest, and engagement with websites and their ads. 
Several works have tried to predict engagement with content in social media based on social interest metrics, such as the number of post comments or likes~\cite{hu2015predicting,claussen2013effects,bakhshi2014faces}. Researchers in information retrieval have also investigated methods to estimate users' satisfaction and engagement with textual and visual Web search engines~\cite{song2013evaluating,jing2015visual,zhang2018well}.
In computational advertising, existing works have tried to improve ad serving based on target engagement metrics~\cite{barbieri2016improving,yi2015dwell}, or to directly predict ad click-through rates~\cite{ling2017model}. 

\xhdr{User engagement on Wikipedia}
Users' reasons for engaging with Wikipedia are varied, and consequently bring about different usage behaviors~\cite{singer_why_2017,lemmerich_why_2019}, ranging from in-depth information seeking to using Wikipedia as a stopover towards other locations on the Web~\cite{teblunthuis_dwelling_2019}.
The most relevant prior work is a study by \citeauthor{piccardi2020quantifying}~\cite{piccardi2020quantifying}, who used the same dataset as the present paper, but focused exclusively on external links in cited references, whereas the present study considers all external links, with a focus on official links listed in infoboxes.
\citeauthor{piccardi2020quantifying}~\cite{piccardi2020quantifying} found that users engage little with the external links included in references, and that they do so more frequently from short, possibly unsatisfactory Wikipedia articles and in order to visit recent content, open access sources, and references about life events (births, deaths, marriages, etc.).
A similar negative relationship between Wikipedia article quality and user engagement with external sources via references was found specifically for medical content~\cite{Maggio797779}.

Wikipedia's own traffic is influenced by its connections to the larger Web ecosystem, and its interdependence with search engines, and Google in particular, has real consequences. On the one hand, Wikipedia's content improves Google search results (e.g., via content snippets); on the other hand, this might keep users from visiting Wikipedia itself~\cite{mcmahon2017substantial}.
In the present paper, we take a novel angle by focusing not on traffic from the rest of the Web to \WP, but on traffic from \WP to the rest of the Web.

\xhdr{The economic value of Wikipedia} 
The value of Wikipedia to the world is not only high, but also difficult, if not impossible, to fully quantify in purely economic terms. 
It has been shown that Wikipedia is essential---or has the potential to be---in a variety of spillovers with substantial economic impact. For instance, it is of critical importance to Web search engines, such as Google~\cite{mcmahon2017substantial}, and has also been shown to be useful to improve, or even predict, financial markets~\cite{tsinghua_university_beijing_impact_2013,moat_quantifying_2013,matrasulov_anticipating_2014}.
Wikipedia can be used to inform economic development policies~\cite{sheehan_predicting_2019}, improve the visibility of places, with direct positive consequences on tourism~\cite{hinnosaar2019wikipedia}, and even predict and monitor global health and diseases~\cite{generous_global_2014,hickmann_forecasting_2015}.
Furthermore, Wikipedia has been shown to influence the very development of science~\cite{thompson_science_2017}.
Nevertheless, and perhaps surprisingly, to the best of our knowledge Wikipedia's economic value as an information gateway to the Web has rarely been discussed in previous work. In a rare exception, researchers have considered the value of Wikipedia in providing traffic to Reddit and Stack Overflow~\cite{vincent_2018}. 


\section{Data and definitions}
\label{sec:Data}

\subsection{\WP client-side logs}

In order to analyze user engagement with \WP's external links, we made use of the dataset collected by \citeauthor{piccardi2020quantifying} \cite{piccardi2020quantifying}.
This dataset consists of logs of all reader interactions with external links in English Wikipedia articles over the one-month period from 24~March to 21~April 2019.
The data was captured by the browser on the client side and includes all clicks on external links and a uniformly random sample (33\%) of all pageview events, organized into \textit{sessions,} \ie, sequences of events from the same user in the same browser tab.
This article reports results
using the full dataset when describing external\hyp click events.
Whenever pageview counts are involved, we extrapolate from the 33\% sample.

The data was collected in accordance with Wikimedia's privacy policy%
\footnote{\url{https://foundation.wikimedia.org/wiki/Privacy_policy}}
and processed exclusively on Wikimedia computing machinery.
Although the data does not contain personally indentifiable information beyond what is implicit in browsing behavior, it cannot be shared publicly.
For transparency, we publish our data analysis code at 
\url{https://github.com/epfl-dlab/WikipediaAsWebGateway}.

\subsection{Article characteristics}
\label{sec:Page characteristics}

At the time of data collection, Wikipedia had around 5.8M articles, which were loaded by readers more than 4.5 billion times during the month studied.
We characterized each article by popularity (pageviews during the month studied) and length (number of characters).
The popularity distribution was very skewed: 50\% of the articles had fewer than 42 views, 90\% had fewer than 894 views. In contrast, the average number of pageviews in one month was 700.
The most visited 1,550 articles, which represented 0.02\% of all articles, accounted for 10\% of all pageviews.
The most visited pages were articles about topics that were trending in April 2019, such as \cpt{Nipsey Hussle} (5.7M views), \cpt{Notre-Dame de Paris} (4.7M), \cpt{Bonnie and Clyde} (3.5M), or \cpt{Game of Thrones (season 8)} (2.6M). Most of the articles were short, and similar to popularity, the number of characters showed a skewed distribution, with a median of 3,888, and an average of 7,793 characters.

\xhdr{ORES topics}
ORES%
\footnote{\url{https://www.mediawiki.org/wiki/ORES}}
is a toolkit offered by Wikimedia that, among other things, includes functionality for labeling articles with topics, based on a manually curated taxonomy of 64 topics~\cite{ORES} derived from WikiProjects.%
\footnote{\url{https://en.wikipedia.org/wiki/Wikipedia:WikiProject}}
Based on this categorization, ORES offers a classifier that predicts, for a given article, its probability of belonging to each of the 64 topics.
Since a single article may belong to multiple topics, the 64 probabilities generally do not sum to 1.
We used topic labels in binarized form, considering an article to belong to a topic if the corresponding probability is greater than 50\%.
Note that, although the taxonomy is hierarchically organized in two levels,
in this work we only considered the 57 lower-level topics (listed along the $x$-axis of \Figref{fig:topics_link_ctrate}).
Having run the classifier on all articles in the dataset, we observed that overall the most common topic was \cpt{Biography} (1.7M articles), followed by \cpt{Sports} (1.4M) and \cpt{North America} (950K). The least common topics were \cpt{Eastern Africa} (11K) and \cpt{Libraries \& Information} (14K).

\subsection{External links}
External links form our central object of study, so we extracted detailed information about them from \WP articles. Since parsing content from articles in wikitext format might result in missing hyperlinks~\cite{Mitrevski_Piccardi_West_2020}, we extracted the external links from the articles in HTML format instead. As this paper focuses specifically on those \WP links that lead to external websites, we adopt the convention that, whenever we simply say ``link'', we implicitly mean ``external link''.

We partitioned the external links in the dataset into three classes, according to their position on the page:
\textit{infobox,} article \textit{body,} and \textit{references}. \textit{Infoboxes} are tables (typically rendered by the browser on the right-hand side of the page on desktop devices, or at the top of the page on mobile devices) that summarize key information by adding semi-structured content (see \Figref{fig:screenshot}).
In addition to images and textual properties, this area can contain---potentially many---external links pointing to external geolocation services, official registries, or official websites.
The links in an article \textit{body} appear inline within the main textual content of the page or in dedicated sections such as ``External links'' or ``See also''.
Links in article bodies are more heterogeneous, including links to social media pages, PDF documents, or related external material.
Finally, we considered as \textit{references} all links used to cite external content in support of a statement. Typically they appear at the bottom of the page, reachable from the article body via numbered link anchors.

During the period considered, Wikipedia had 5.3M articles that contained at least one of 63.1M external links (totaling 49.8M unique target URLs).
\Tabref{tab:general_stats} (column ``Total links'') summarizes these values.
In total, 35.3M (56.0\%) of these links appeared in references, 24.9M (39.5\%) in article bodies, and 2.8M (4.5\%) in infoboxes.
Around 1.3M articles in English Wikipedia had an infobox with links, and the average number of links per infobox in these articles was 2.08. Links spanned from official company links (e.g., \url{schlenkerla.de}) to geocoordinates on \url{geohack.toolforge.org} to institutional registries (\eg, National Register of Historic Places).

\subsection{Official links}
\label{sec:Official links}

To further qualify the infobox links, we designed a binary classifier that can distinguish between \textit{official links} and other types of link.
It was trained on a random sample of 2,000 infobox links, manually annotated as ``official'' or ``other''.
This resulted in a training set of 387 official links and 1,613 other links. To characterize each link, we then computed the following features:
\begin{itemize}
  \item \textbf{URL length:} Number of characters in the URL path (guided by the intuition that official\hyp link URLs tend to be short).
   \item \textbf{Similarity of URL with article title:}
    Motivated by the usefulness of character $n$-grams for URL-based topic classification~\cite{Baykan}, we computed the character $n$-grams ($n=1,\dots,4$) of the title of the article where the link was placed, the link's anchor text (if non-empty), the domain of the link URL, and the path of the link URL. We then computed, and used as features, the Jaccard similarity of sets of $n$-grams for three pairs: title\slash URL-domain, anchor-text\slash URL-domain, title\slash URL-path.
 \item \textbf{Similarity of context with marker words:} Jaccard similarities between the character $n$-gram sets ($n=1,\dots,4$) of high\hyp precision marker words (``official'', ``website'', ``homepage'', ``URL'') and of the link's anchor text and context (\ie, the text within the same \texttt{<TR>} tag as the link).
\end{itemize}

We used these 10 features and the manual labels in a random forest classifier, achieving 5-fold cross\hyp validated precision 0.980 (SD $0.009$), recall 0.983 (SD $0.005$), and F1 score 0.982 (SD $0.007$).

Applying this classifier to all the links in the dataset, we found that 506K of the 63.1M external links corresponded to the official website of the entity described in the respective article. 
Broken down by article topic, the largest number of official links was associated with
\cpt{North America} 
(27.2\%), \cpt{Europe} 
(25.0\%), \cpt{Media} 
(20.2\%), \cpt{Biography} 
(18.9\%), \cpt{Asia} 
(17.1\%), \cpt{Education} 
(10.5\%), and \cpt{Business and economics} 
(9.5\%),
with the latter containing mainly articles about companies.
(The percentages sum to more than 100\% because articles may belong to multiple topics.)

As expected, according to the classifier,
the vast majority (98.1\%) of articles with an official link had exactly one official link,
and conversely,
the vast majority (97.1\%) of official links appeared as such in exactly one article's infobox.
This has the added advantage that official links can be characterized by features derived from their corresponding \WP articles (\eg, topics, content words).


\subsection{Definitions}
\label{sec:defs}

\xhdr{Definition: \ctr (CTR)}
Our main metric for measuring engagement with external links is the \textit{\ctr (CTR)},
which, intuitively, is simply the number of times a link was clicked, divided by the number of times the link was displayed (by virtue of being contained in an article that contained the link).
In practice, care must be taken, as it frequently happens that the same article is viewed multiple times in the same session, \eg, because the user refreshes the page or clicks the back button.
To guard against overcounting such multiple views, we grouped all pageviews of the same article $a$ that occurred during the same session $s$ and call the unique pair $(a,s)$ one visit of $a$.

With $N_l$ as the number of visits (\ie, distinct $(a,s)$ pairs) upon which link $l$ was displayed,
and $C_l$ as the number of visits upon which link $l$ was clicked,
we define the CTR of link $l$ as
$C_l/N_l$,
\ie, the fraction of visits upon which $l$ was clicked, out of all visits upon which $l$ was displayed.
Since each official link belongs to exactly one article (with extremely rare exceptions; \cf\ \Secref{sec:Official links}), we may also, in a slight abuse of terminology, speak of the ``CTR of an article'', implying the CTR of the official link associated with the article.

In order to reliably estimate CTRs, we need to avoid small denominators, so we restricted our analyses to links that were displayed upon at least 300 pageview events.
In the case of official links, this resulted in a set of 160K links (and their corresponding articles).

\xhdr{Definition: click time}
In order to capture how long users dwell on an article before they click an external link, we define the notion of \textit{click time,}
which measures the number of seconds between the pageview event on which the link was displayed and the click on the link itself.
If the same external link was clicked multiple times in the same session, we only considered the first pageview that was accompanied by an external click.

Since click times are unbounded above and follow a heavy-tailed distribution, we used medians, rather than means, for aggregation.

\begin{figure}[t]
    \begin{minipage}[t]{.49\columnwidth}
        \centering
        \includegraphics[height=3cm]{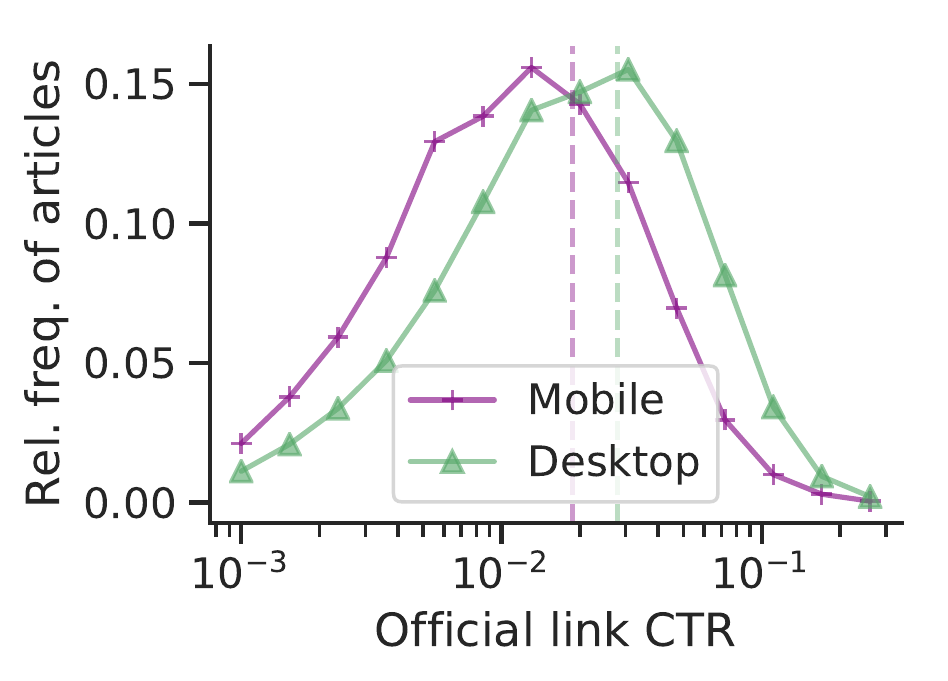}
        \subcaption{\Ctr}\label{fig:ctr_by_device}
    \end{minipage}
    \hfill
    \begin{minipage}[t]{.5\columnwidth}
        \centering
        \includegraphics[height=3cm]{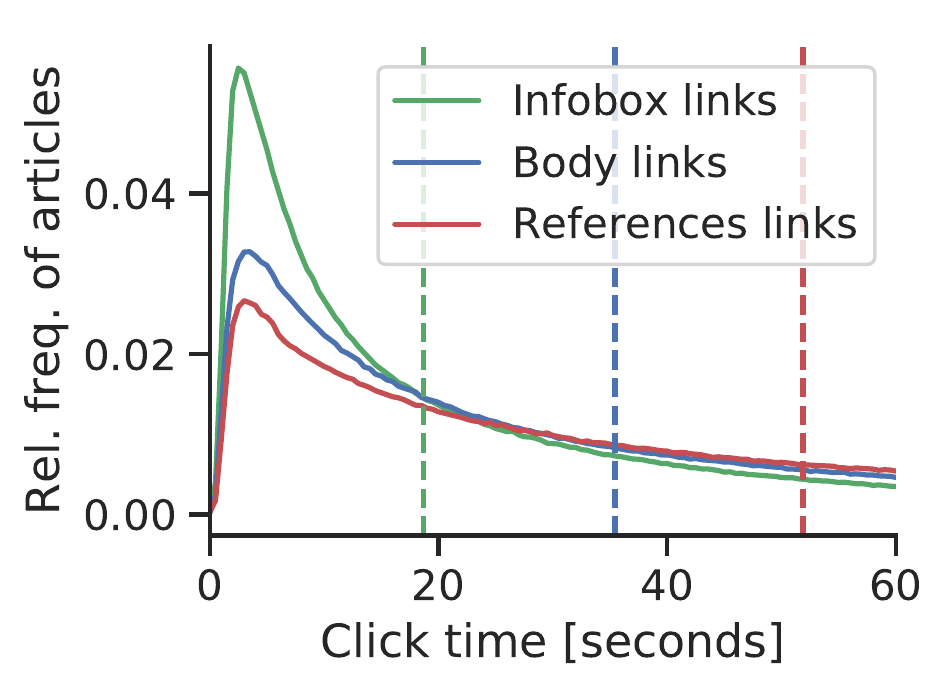}
        \subcaption{Click time}\label{fig:time_by_link_type}
    \end{minipage}
    \vspace{-3mm}
\caption{
Usage of external links.
(a)~Distribution of \ctr of official links by device type (vertical lines: means).
(b)~Distribution of click time by link type (vertical lines: medians).
}
\label{fig:ctr_and_time}
\end{figure}


\section{RQ1: Level of engagement with external links}
\label{sec:RQ1}

We start our analysis by quantifying the level of engagement with external links, both overall (\Secref{sec:Overall click statistics}) and by article topic (\Secref{sec:Click-through rates by topic}).

\begin{figure*}[t]
  \includegraphics[width=0.90\textwidth]{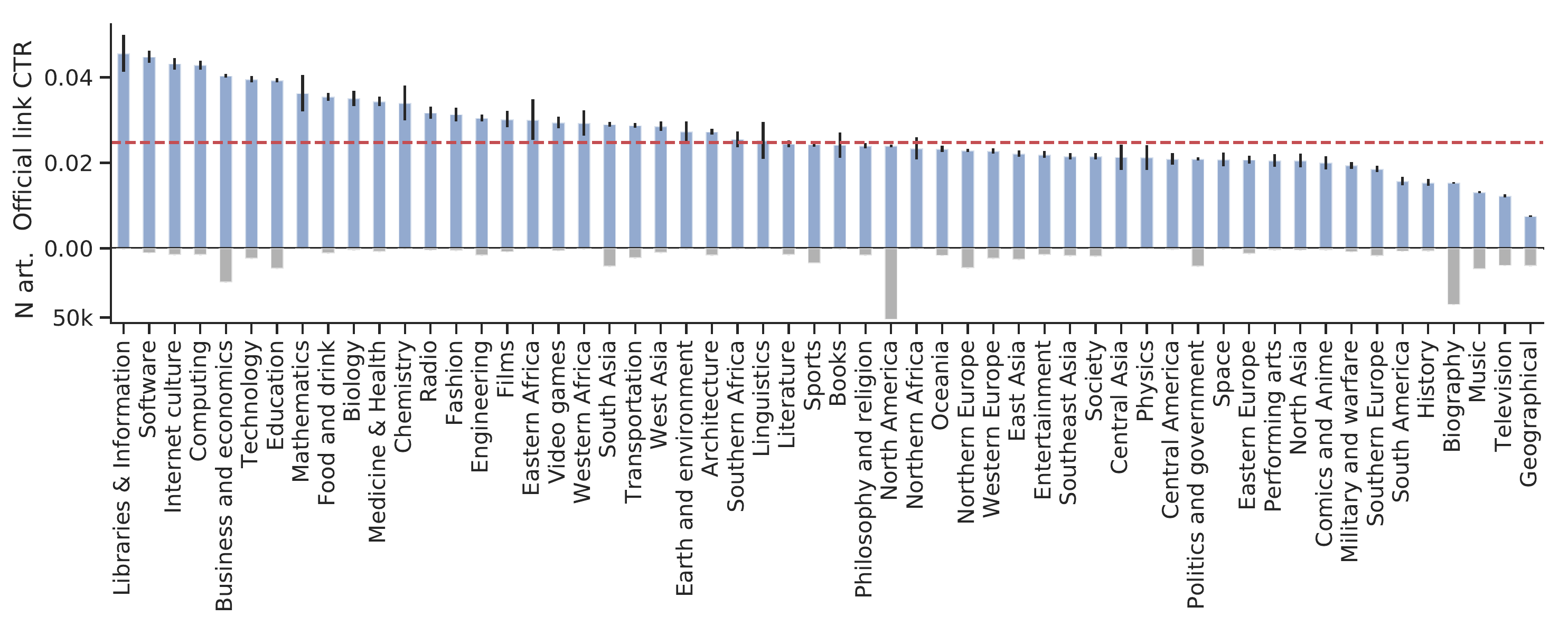}
    \vspace{-3mm}
  \caption{Official-link \ctr by article topic.
  \textit{Blue bars:} means with bootstrapped 95\% confidence intervals.
  \textit{Gray bars:} number of articles with official links.
  \textit{Red dashed line:} global mean.}
  \label{fig:topics_link_ctrate}
\end{figure*}

\subsection{Overall click statistics}
\label{sec:Overall click statistics}

Overall click statistics are summarized in \Tabref{tab:general_stats}.
During our one-month data collection period, there was a total of around 4.5 billion \WP pageviews, which led to around 43.1M clicks on external links.
The total volume of external clicks was distributed roughly evenly over the three classes of external links:
those in infoboxes (12.5M),
those in references (14.2M),
and those in article bodies (16.2M).
As the vast majority of external links was located in references (56.0\%) and article bodies (39.5\%),
the CTR of infobox links (0.90\%) vastly exceeded that of links in references (0.03\%) and article bodies (0.14\%).
To ascertain that this was not simply caused by the fact that infobox links appear higher up on the page, we also computed the CTR of article\hyp body links appearing in the top 20\% of the page.
This yielded a CTR of 0.20\%, much closer to the 0.14\% of article\hyp body links overall than to the 0.90\% of infobox links.

\xhdr{Official links}
Official links play a key role.
Although they constituted only 18\% of the 2.8M infobox links, they accounted for 78\% of the 12.5M clicks on infobox links,
with a CTR of 2.47\%, nearly 3 times as high as that of infobox links overall.
The average CTR was even higher on desktop devices, where it reached 2.78\%, \vs\ 1.87\% on mobile
(\Figref{fig:ctr_by_device}).
Given their prominence, we shall focus mostly on official links from here on, and
unless stated otherwise, we henceforth refer to official links when simply writing ``links''.

\xhdr{Geographical differences}
The top 5 countries by pageview volume were, in this order, the United States,
the United Kingdom, India, Canada, and Australia.
They generated 71.6\% of the total traffic.
Among these countries, the CTR on official links was highest in the U.S.\ (2.36\%), followed by India (2.14\%), the U.K.\ (1.53\%), Canada (1.38\%), and Australia (1.11\%).

\subsection{Click-through rates by topic}
\label{sec:Click-through rates by topic}

Next, we aim to understand how the \ctr{}s of official links vary by topic as defined by the ORES classifier introduced in \Secref{sec:Page characteristics}.
Since, in the vast majority of cases, official links are associated with exactly one article, we may label each official link with the topics of that article.
(Throughout the discussion that follows, keep in mind that each article, and thus each official link, may be labeled with multiple topics.)

\Figref{fig:topics_link_ctrate} visualizes the mean CTR (in blue),
as well as the number of articles with official links (in gray), by topic.
We see that official links relating to \cpt{Libraries \& information}, \cpt{Software}, and \cpt{Internet culture} had the highest \ctr{}s,
whereas geographical content, media-related content, and biographies on average saw the lowest engagement.

\begin{figure}[t]
    \begin{minipage}[t]{\columnwidth}
        \centering
        \includegraphics[width=0.99\columnwidth]{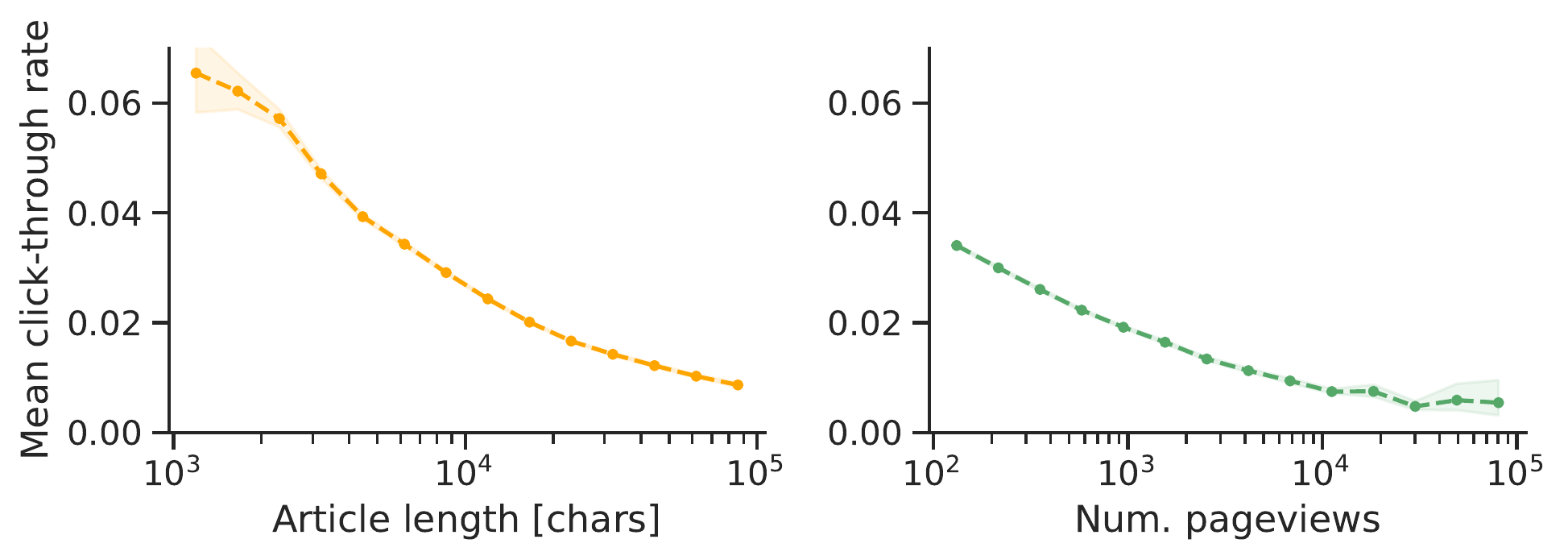}
        \subcaption{\Ctr}\label{fig:ctrate_popularity_length}
    \end{minipage}
    
    \begin{minipage}[t]{\columnwidth}
        \centering
        \includegraphics[width=0.99\columnwidth]{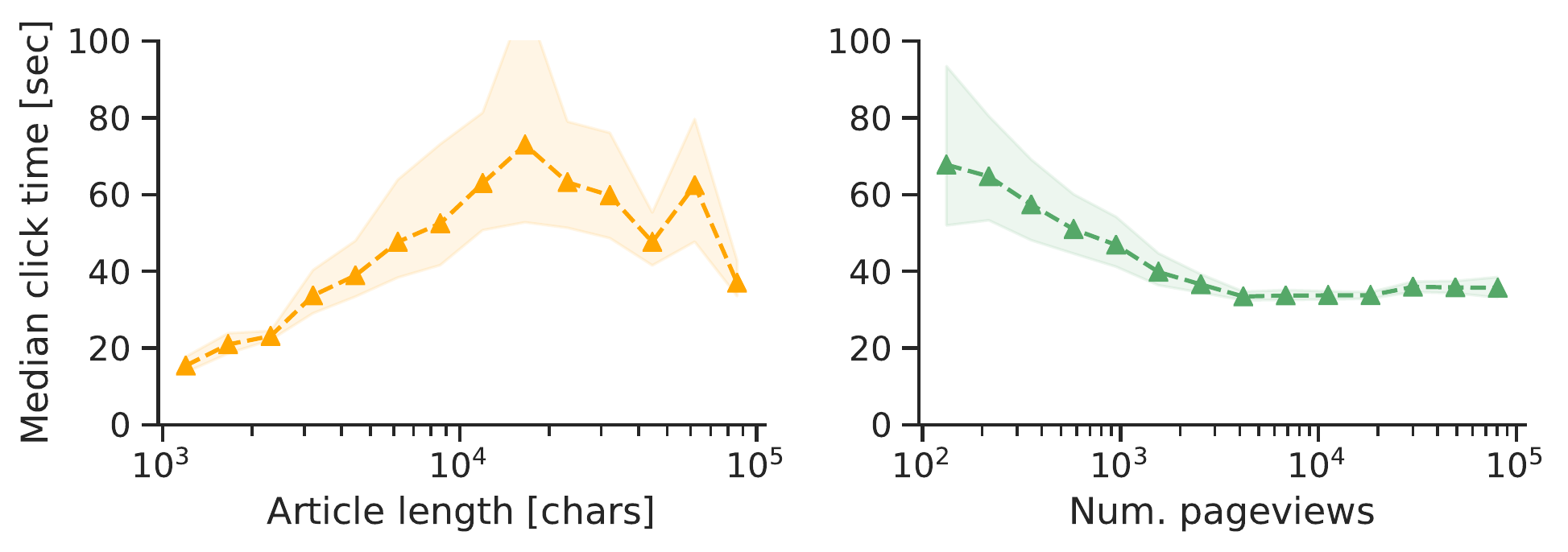}
        \subcaption{Click time}\label{fig:time_popularity_length}
    \end{minipage}
    \vspace{-3mm}
\caption{
(a)~\Ctr and
(b)~click time
of official links as functions of article length (left) and popularity (right),
with 95\% CIs.
Official links on longer pages are clicked more rarely and more slowly;
those on more popular pages are clicked more rarely and more quickly.
}
\label{fig:ctrate_time_popularity_length}
\end{figure}

\xhdr{Controlling for article length and popularity}
The length and popularity of a \WP article correlated strongly and negatively with the CTR of the official link contained in its infobox (\Figref{fig:ctrate_popularity_length}),
possibly because longer articles, by offering more information, reduce the user's need to gather additional information from external links, and because more popular articles are more likely to appear in shallower information\hyp seeking sessions~\cite{piccardi2020quantifying}.
Since length and popularity also vary by topic, they might act as confounds that could potentially explain
an observed variation of CTR by topic.

To tease these two confounds apart from the impact of topics alone, we controlled for length and popularity in a matched analysis, as follows.
We split the set of articles at the median CTR into high- \vs\ low-CTR articles,
and we split the length and popularity ranges into 1,000 equally sized bins each.
We then defined a bipartite graph with edges between articles that fell in different CTR halves, but in the same bins with respect to length and popularity.
Using the Euclidean distance in the space defined by the logarithmic length and popularity as edge weights, we found a minimum matching and retained only the
112K matched articles (out of originally 160K).
This procedure successfully balanced the dataset.%
\footnote{The standardized mean differences in logarithmic length and logarithmic popularity dropped from 0.7 to 0.00017, and from 0.54 to 0.000005, respectively.}

In the balanced dataset, we binarized the CTR by splitting at the median and fit a logistic regression to model whether an article belonged to the high- or low-CTR group, with topic indicators as predictors (pseudo $R^2 = 0.20$, $p < 10^{-307}$).
(The advantage of performing regression modeling rather than a simple comparison of per-topic average CTRs is that topics are correlated, which is accounted for by the regression model.)

\begin{figure}[t]
    \begin{minipage}[t]{\columnwidth}
        \centering
        \includegraphics[width=0.99\columnwidth]{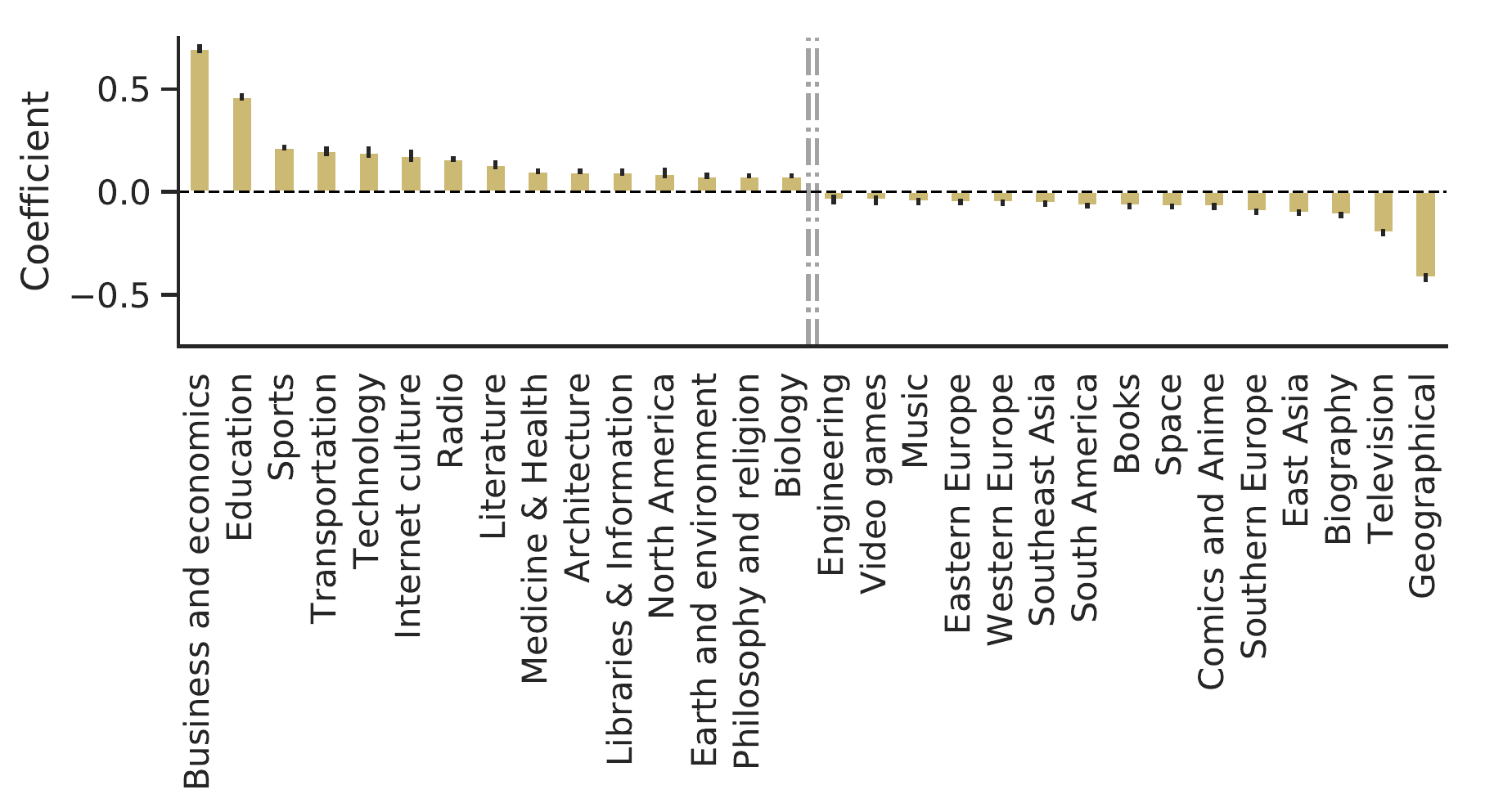}
        \subcaption{Topic-based analysis}\label{fig:coefficients_ctr}
    \end{minipage}
    \\
    \begin{minipage}[t]{\columnwidth}
        \centering
        \includegraphics[width=0.94\columnwidth]{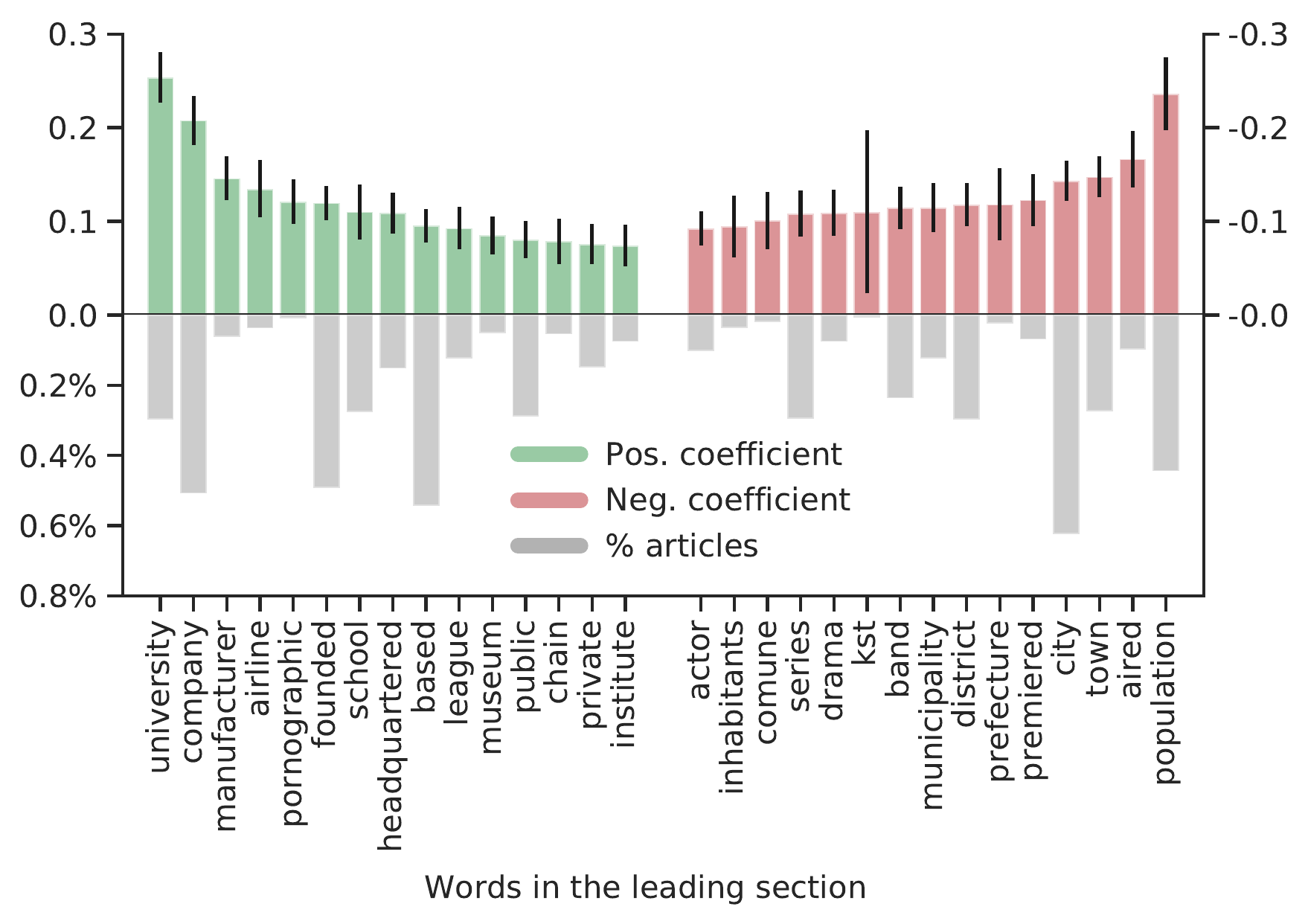}
        \subcaption{Word-based analysis}\label{fig:linguistic_click}
    \end{minipage}
    \vspace{-3mm}
\caption{
Association of \ctr of official links with article properties, captured via 15 largest positive and negative coefficients
(with 95\% CIs)
from binary logistic regression models that predict above- \vs below-median CTR, using as predictors
(a)~article topics
or
(b)~words from lead paragraphs
(controlling for article length and popularity).
Gray bars in~(b): percentage of articles whose lead paragraph contains the word.
}
\end{figure}

The 15 largest positive and negative coefficients, plotted in \Figref{fig:coefficients_ctr}, revealed a slightly different ranking than \Figref{fig:topics_link_ctrate}, with \cpt{Business and Economics} and \cpt{Education} emerging as the strongest predictors of a high CTR,
whereas \cpt{Geographical} and \cpt{Television} remained the strongest predictors of a low CTR.

\begin{figure*}[t]
  \includegraphics[width=\textwidth]{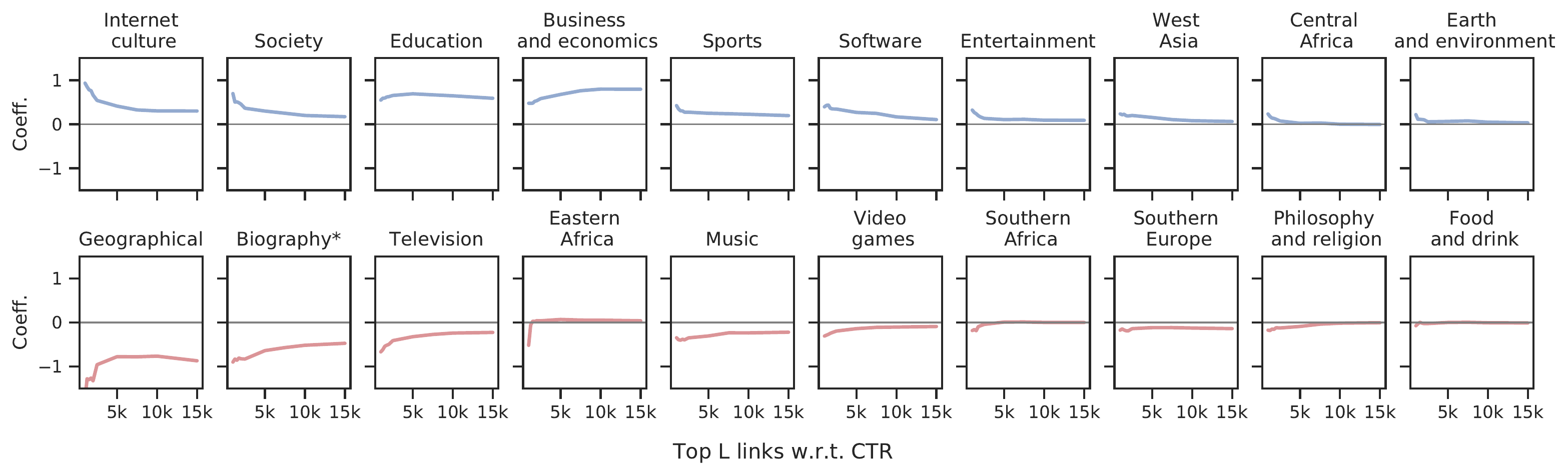}
  \vspace{-7mm}
  \caption{
  Prevalence of topics among most frequently clicked official links.
  We fitted binary logistic regression models that used article topics as predictors to predict if an article's official link is among the top $L$ highest-CTR links.
  Plots show regression coefficients for individual topics (predictors) as functions of $L$ (for values of $L$ between 1K and 15K).
  Topics are sorted by the leftmost values of their curves.
  Sharply decreasing [increasing] curves correspond to topics that are particularly over\hyp represented [under\hyp represented] among the links with the most extreme CTR.
  (More details: \Secref{sec:Click-through rates by topic}, ``Top of the CTR ranking''.)
  }
  \label{fig:topics_ctrate_topk}
\end{figure*}

\xhdr{Top of the CTR ranking}
While manually screening the data, we realized that, among the articles with the highest official\hyp link CTR, there was a disproportionate fraction of articles about websites (which are generally classified by ORES under the topic \cpt{Internet culture}),
and in particular websites related to file sharing and pornography, some with CTRs of 40\% or more,
\eg, Library Genesis (47\%), RARBG (45\%), or The Pirate Bay (43\%).
To determine whether official links of \WP articles about websites dominated the top of the CTR ranking in general, we repeated the above regression analysis with a small modification:
instead of predicting the top half \vs\ the bottom half of the article ranking with respect to CTR, we now predicted the top $L$ articles (an absolute, rather than relative number) \vs the same number of samples from the bottom half, matched on length and popularity.
This way, plotting the fitted coefficients for a given topic as a function of $L$ reveals whether the topic is particularly over\hyp represented among the highest-CTR official links (manifested in a sharply decreasing curve).
The results, presented in \Figref{fig:topics_ctrate_topk}, clearly show that \cpt{Internet culture}---a topic held by most articles about websites---is indeed particularly over\hyp represented among the articles with the very highest official\hyp link CTR.
Similar effects were observed for
\cpt{Society} (a loose mix of articles),
\cpt{Sports},
\cpt{Software}, and
\cpt{Entertainment},
among others.
On the contrary, we observed that
\cpt{Geographical},
\cpt{Biography}, and
\cpt{Television},
among others,
were particularly under\hyp represented among the highest-CTR official links.

\xhdr{Fine-grained topical analysis}
The topics from the ORES classifier used above are rather broad.
In order to obtain more fine-grained insights,
we conducted a word- rather than topic-level analysis, where we represented an official link by the words contained in the lead paragraph of the article in whose infobox the link appeared as an official link (via $z$-score\hyp standardized TF-IDF vectors restricted to the 3,000 most frequent words across all articles with official links).
Mirroring the above regression analysis, but now using words rather than topics as predictors (pseudo $R^2=0.31$, $p < 10^{-307}$), this analysis revealed words associated with high- \vs\ low-CTR links.
\Figref{fig:linguistic_click}, which shows the 15 words with the largest positive and negative coefficients,
confirms our previous findings while adding nuance.
We see that \cpt{Education} specifically marks high-CTR links about universities, schools, institutes, and museums;
\cpt{Business and economics}, about companies, manufacturers, chains, and airlines;
and \cpt{Internet culture}, about adult websites.

\xhdr{Summary}
Taking stock of the findings so far,
we reiterate that official links play a key role among \WP's external links, with CTRs far above those of other types of external link.
We observed a large amount of variation depending on article topics, with official links associated with articles about websites, software, businesses, education, and sports seeing particularly high engagement.


\section{RQ2: Patterns of engagement with external links}
\label{sec:RQ2}

Above, we established which types of external link have a particularly high CTR.
Next, we investigate more closely the patterns by which users engage with external links.

\subsection{Click time}
\label{sec:Click time}

We start by analyzing the click time (\cf\ \Secref{sec:defs}), which captures how long users dwell on an article before leaving it toward an external website via an external link.

Click time statistics are summarized in \Tabref{tab:general_stats}, and click time distributions are plotted in \Figref{fig:time_by_link_type},
for the three types of link:
those appearing in infoboxes, article bodies, and references, respectively.
(Note that, although we consider only official links in most of this paper---and indeed in the rest of this section---we nevertheless report this basic statistic for all types of external link beyond official links only.)
The global median click time was 32.9 seconds (31.8 seconds for desktop, 34.4 seconds for mobile),
with a much lower value for infobox links (18.7 seconds; 20.1 seconds for official links),
and larger values for the article-body links (35.4 seconds)
and reference links (51.8 seconds).
The short click time of infobox links, however, seems to be due to their prominent position within articles: when approximately controlling for position by considering only article-body links in the top 20\% of the page, the median click time dropped to 22.2 seconds, only 10\% longer than for infobox links.

After this general characterization, we from here on focus on official links in infoboxes.
Similar to CTR (\Figref{fig:ctrate_popularity_length}),
the click time of official links was correlated with article length and popularity (\Figref{fig:time_popularity_length}), such that
clicks took more time on longer and on less popular articles.
When analyzing click time by topic, we thus again controlled for these two factors via matching, as they could otherwise confound the analysis.
Analogously to the setup of \Secref{sec:Click-through rates by topic},
we split the set of articles at the median click time into articles with ``slow'' \vs\ ``fast'' clicks on official links, and subsequently found a bipartite matching in order to ensure that the distribution of length and popularity was nearly identical in articles with ``slow'' \vs\ ``fast'' official\hyp link clicks.%
\footnote{The standardized mean differences in logarithmic article length and logarithmic popularity dropped
from 0.076 to 0.0001,
and from 0.30 to 0.00005,
respectively.
138K of the original 160K data points were retained after matching.
}
We then fitted a linear regression on the resulting balanced dataset in order to model the logarithmic click time as the outcome variable, using (as in \Secref{sec:Click-through rates by topic}) topic indicators as predictors ($R^2=0.090$, $p < 10^{-307}$).

\begin{figure}[t]
    \begin{minipage}[t]{\columnwidth}
        \centering
        \includegraphics[width=0.99\columnwidth]{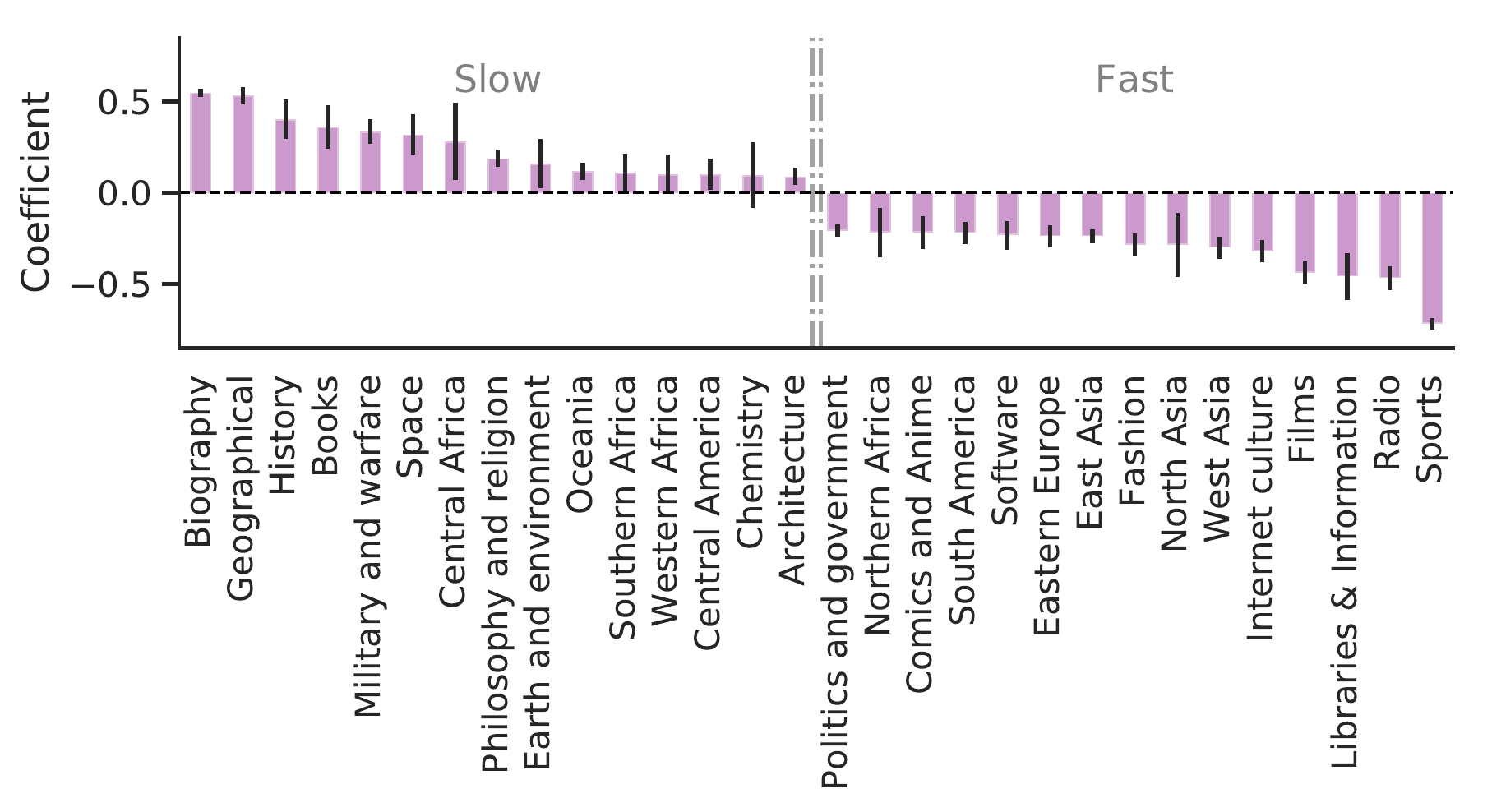}
        \subcaption{Topic-based analysis}\label{fig:coefficients_speed}
    \end{minipage}
    \\
    \begin{minipage}[t]{\columnwidth}
        \centering
        \includegraphics[width=0.94\columnwidth]{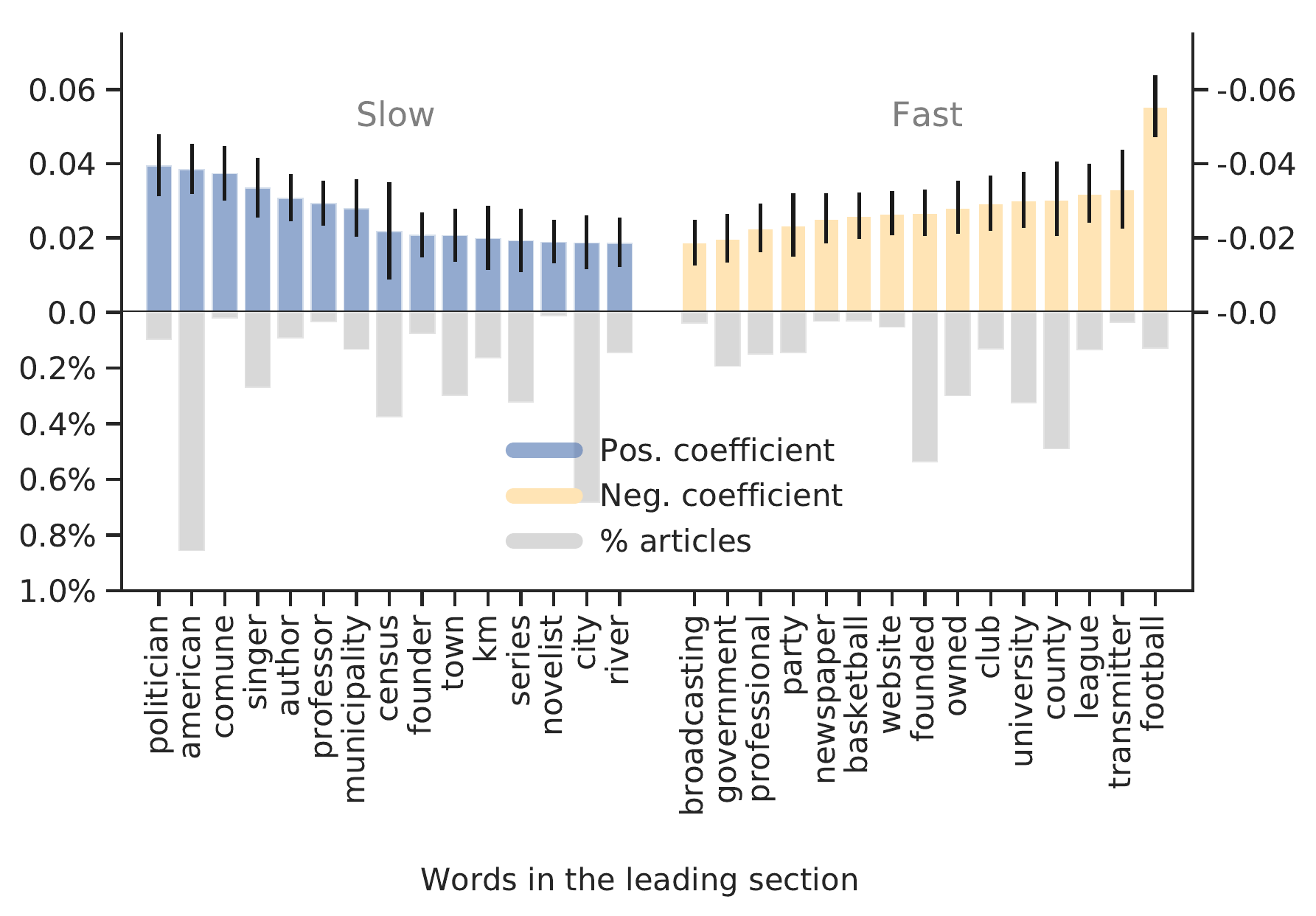}
        \subcaption{Word-based analysis}\label{fig:linguistic_speed}
    \end{minipage}
    \vspace{-3mm}
\caption{
Association of click time of official links with article properties, captured via 15 largest positive and negative coefficients
(with 95\% CIs)
from linear regression models that predict logarithmic click time, using as predictors
(a)~article topics
or
(b)~words from lead paragraphs
(controlling for article length and popularity).
Gray bars in~(b): percentage of articles whose lead paragraph contains the word.
}
\end{figure}

The 15 largest positive and negative coefficients are plotted in \Figref{fig:coefficients_speed}, revealing that clicks on official links to entertainment\hyp related websites occurred faster,
whereas links to websites on more classic encyclopedic topics, such as biographies, geographical content, history, \etc, occurred more slowly.

To gain more granular insights than at the coarse topic level, we again ran an analysis at the word level, parallel to the word-level analysis of \Secref{sec:Click-through rates by topic}, but this time in the linear regression setup just described ($R^2=0.17$, $p < 10^{-307}$).
The words most indicative of slow and fast clicks (\Figref{fig:linguistic_speed})
mirror the findings from the topic-level analysis (\Figref{fig:coefficients_speed}), but we now also see that official links on articles about websites and universities were clicked particularly fast.

\subsection{Wikipedia as a stepping stone}
\label{sec:Wikipedia as stepping stone}

\begin{figure}[t]
    \centering
    \includegraphics[width=0.8\columnwidth]{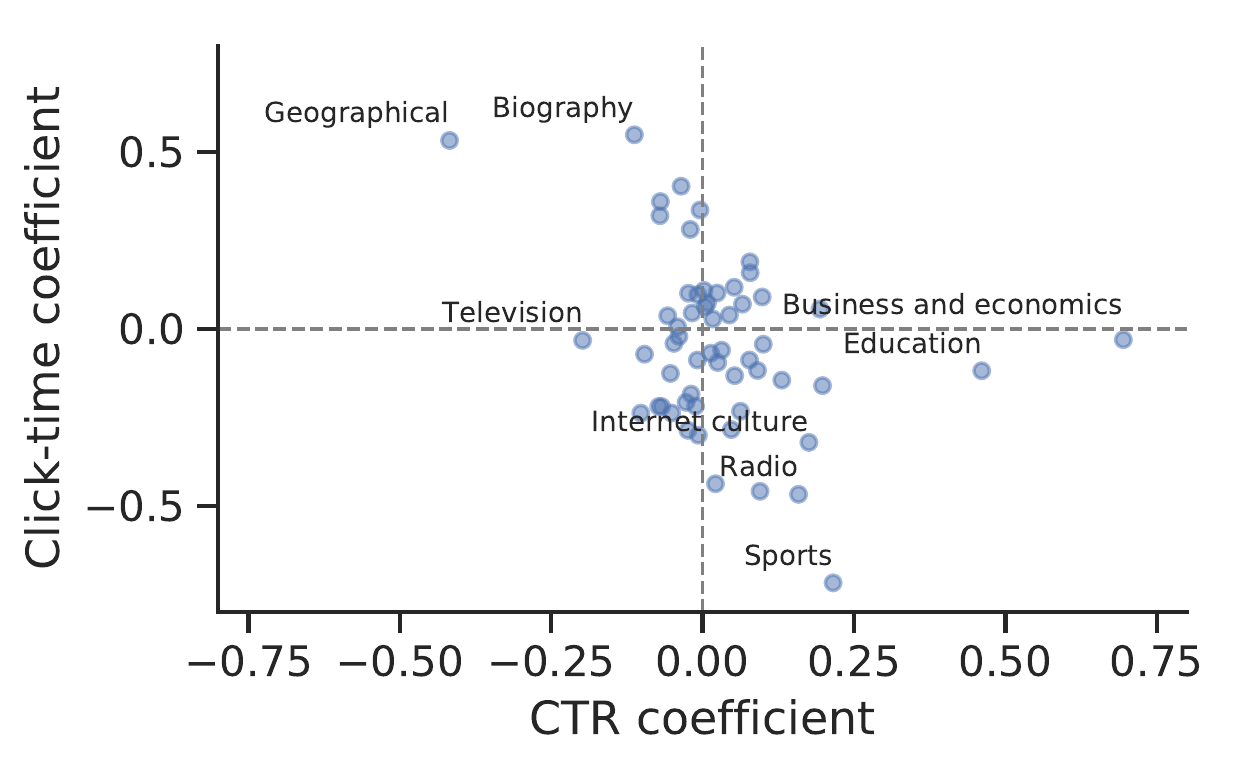}
    \vspace{-3mm}
    \caption{
    \Ctr ($x$-axis) \vs\ click time ($y$-axis) of official links.
    Each point represents one topic.
    CTR and click time of a topic captured in terms of the topic's coefficient in the regressions summarized in \Figref{fig:coefficients_ctr} and \ref{fig:coefficients_speed}, respectively.
    }
    \label{fig:coeff_ctrate_vs_time}
\end{figure}

We just noted that official links on articles about websites typically have short click times.
Moreover, in \Secref{sec:Click-through rates by topic} we had observed that such links are also strongly over\hyp represented among the highest-CTR links.
This observations led us to hypothesize that the main interest of users interacting with these links might be to find these links to begin with, rather than to find the content that surrounds the links in their respective \WP articles.
In other words, we hypothesized that \WP serves as a ``stepping stone'' toward external websites, whereupon users barely set foot onto \WP before leaving again towards different content that they actually were intent on finding.
To further investigate this hypothesis, we considered the relationship between CTR and click time, visualized in \Figref{fig:coeff_ctrate_vs_time}, which presents a scatter plot of the coefficients obtained from the previously described regressions for CTR ($x$-axis; \cf\ \Secref{sec:Click-through rates by topic}) and click time ($y$-axis; \cf\ \Secref{sec:Click time}).
As the plot shows, topics with a high CTR tended to have low click times (lower right quadrant),
notably \cpt{Sports}, \cpt{Radio}, and \cpt{Internet culture} (a topic that, as mentioned, primarily tags articles about websites).
We take this as another indicator of the existence of a class of articles from which users leave \WP frequently and fast.


If indeed a distinct class of articles on \WP are used heavily as stepping stones, then we should be able to identify many articles that were visited primarily from outside of \WP (\eg, from search engines), rather than from other \WP articles (via internal navigation), and that had a high CTR on the external links they contain.
With the goal of finding such articles, we define the \textit{external\hyp referrer frequency (ERF)} of an article as the fraction of all visits made to the article via a click from a referral page external to \WP.
Note that external referrals almost exclusively stemmed from search engines:
about 70\% of them specified a search engine referrer URL in the logs,
and about 29\% did not specify any referrer URL, but it is suspected that a majority of these visits also come from search \cite{searchengineland2014}.
Hence, we included empty referrers in our analysis, although the conclusions were identical when excluding them.

\begin{figure*}[t]
    \begin{minipage}[t]{0.24\textwidth}
        \centering
        \includegraphics[width=\textwidth]{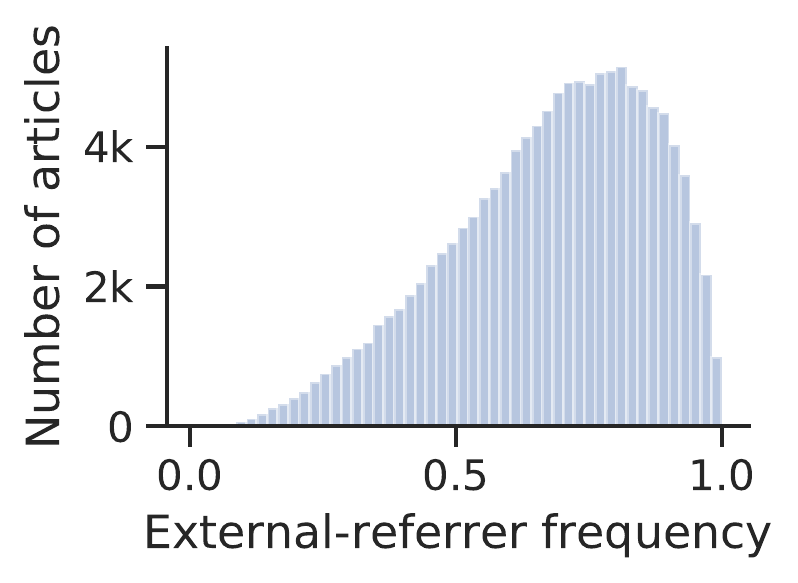}
        \subcaption{}\label{fig:ext_ref_total_pages}
    \end{minipage}
    \hfill
    \begin{minipage}[t]{0.24\textwidth}
        \centering
        \includegraphics[width=\textwidth]{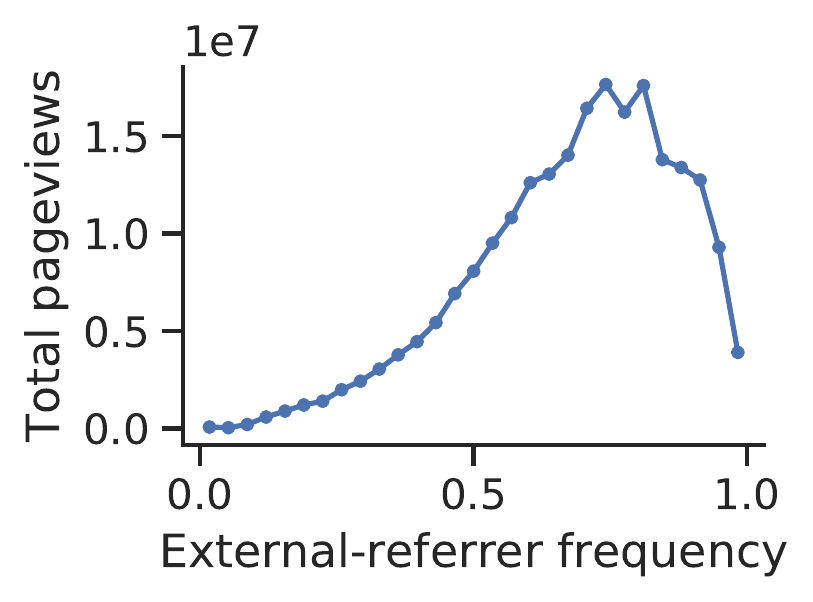}
        \subcaption{}\label{fig:ext_ref_pageviews}
    \end{minipage}
        \hfill
    \begin{minipage}[t]{0.24\textwidth}
        \centering
        \includegraphics[width=\textwidth]{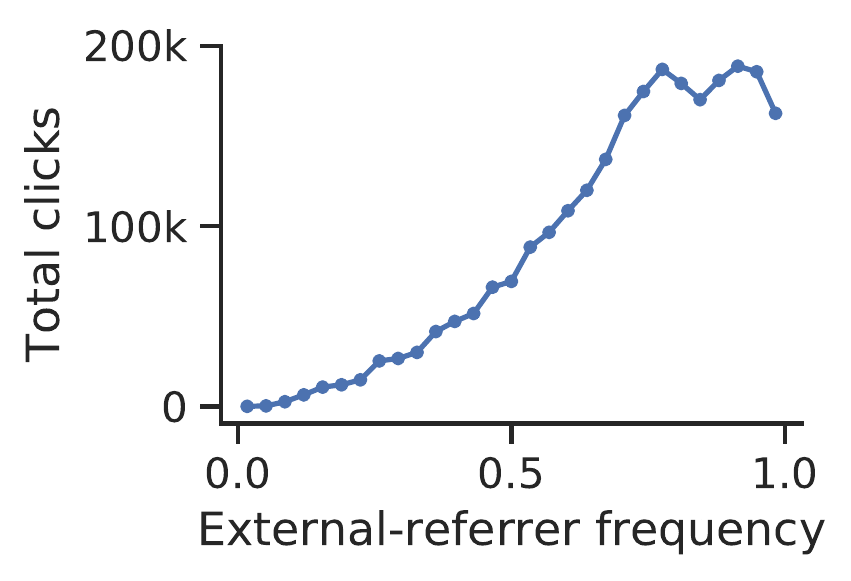}
        \subcaption{}\label{fig:ext_ref_clicks}
    \end{minipage}
        \hfill
    \begin{minipage}[t]{0.24\textwidth}
        \centering
        \includegraphics[width=\textwidth]{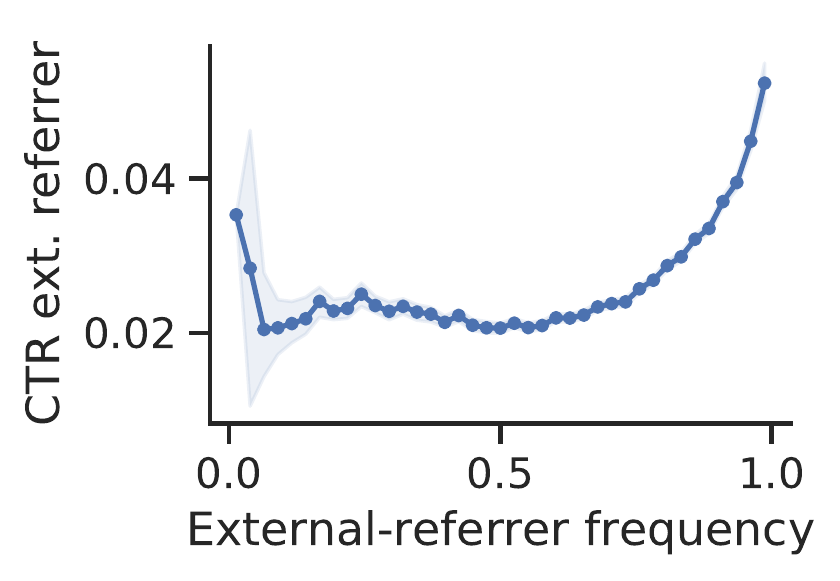}
        \subcaption{}\label{fig:ext_ref_lctr}
    \end{minipage}
    \vspace{-3mm}
\caption{
    Quantification of \WP's role as a stepping stone toward external websites.
    (a)~Histogram of \textit{external\hyp referrer frequency (ERF)} of \WP articles, where ERF is defined as the fraction of times the article was visited via a referral page external to \WP.
    (b)~Total number of pageviews of articles within each ERF bin.
    (c)~Total number of official-link clicks of articles within each ERF bin.
    (d)~Official-link CTR upon pageviews with an external referrer (most likely a search engine), with 95\% CIs.
    Articles with an extreme ERF close to~1 are rare (a), but generate a disproportionately large number of official-link clicks (c~\vs~b), especially when reached from search engines (d).
}
\label{fig:external_internal}
\end{figure*}

The ERF histogram, plotted in \Figref{fig:ext_ref_total_pages},
shows that most articles were primarily visited from outside of \WP, but only few articles had a very high ERF close to~1.
Although extreme-ERF articles were in total visited less than medium-ERF articles (\Figref{fig:ext_ref_pageviews}),
they generated a total number of external clicks comparable to that generated by medium-ERF articles (\Figref{fig:ext_ref_clicks}).
The most important piece of evidence, however, comes from \Figref{fig:ext_ref_lctr}, which plots, on the $y$-axis, the mean CTR for situations where the respective article was reached from an external referrer.
The plot shows that the articles that were nearly exclusively reached from external referrers (with an ERF close to 1) are precisely those articles that also had the highest CTR after being reached from an external referrer.%
\footnote{
This analysis only considered articles with at least 100 visits from external referrers, in order to avoid noise due to division by small numbers.
}

Taken together, these facts provide evidence of a class of articles that serve as mere ``stepping stones'', ``revolving doors'', or ``in-and-outs'':
users come from elsewhere in order to find a particular link and immediately leave \WP by clicking that link.

But why, then, would users go through \WP in the first place, if all they want is to go to a website linked from \WP? We shall discuss potential reasons for this behavior in \Secref{sec:Discussion}.

\section{RQ3: Economic value of external links}
\label{sec:RQ3}

Our final set of analyses aims to estimate the monetary value of the traffic that \WP drives to external websites. 

The idea behind our calculations is the following.
Search engines generally charge website owners money in exchange for driving clicks to their sites from sponsored search results (\eg, Google Ads), whereas \WP conveys a large volume of traffic to official websites for free. 
We therefore ask,
``How much money would a search engine want from website owners to obtain, via ads, the same number of clicks they obtain from \WP for free?''

While we could be asking, ``How much money could Wikipedia earn by charging a fee for external clicks?'',
we consider this counterfactual scenario too far from reality:
\WP is open and free by design, and it functions rather differently from platforms driven mostly by advertising.
Moreover, we could not estimate the ``price paid'' to \WP in this scenario, as we do not know whether website owners would be willing or able to actually pay for any hypothetical click fees.
On the contrary, estimating the ``price asked''---the cost of online ads---is entirely feasible.

\subsection{Methodology}
\label{sec:RQ3 methodology}

\xhdr{Google Ads}
To estimate the price of achieving a certain set of URL clicks via online ads, we used the Google Ads API, with ``sponsored search'' as the ads network.
Google Ads is one of the most prolific advertising platforms, and the primary source of revenue for Google's parent company, Alphabet \cite{AlphabetRevenue}.
The Google Ads API allows advertisers to create campaigns for promoting a URL by placing bids on campaign\hyp related search keywords.
The bid expresses the maximum amount that the website owner is willing to pay each time the promoted URL is clicked when shown on the search result page for the respective keyword.
When a user searches for a keyword specified by the campaign, an auction system determines which sponsored URL to show among all the candidates competing for the keyword.%
\footnote{
\label{fn:broad search}
With the ``broad search'' option, the number of matching searches increases by also considering substrings and permutations of the tokens in campaign keywords.
}
When the user clicks a promoted link, the campaign owner pays the auction value.
Note that the paid price is not necessarily equal to, but only bounded by, the website owner's bid.

\xhdr{From URLs to keywords}
Our intended analysis started from clicks on official links (URLs) observed in the \WP logs
and aimed to estimate how much these clicks would cost when obtained via Google Ads instead of \WP.
The Google Ads API, on the contrary, requires keywords, not URLs, as input.
Thus, in order to leverage Google Ads for our analysis, we had to work our way backwards and determine appropriate keywords that a website owner might use to advertise a given URL.
Since the choice of the right keywords is critical to increase a website's discoverability while keeping ad costs down~\cite{SCHOLZ201996}, Google Ads offers a tool called \textit{Keyword Planner,} which, given a website URL, generates a set of relevant keywords, alongside information on the historical bidding range and search volume for those keywords.
Using the Keyword Planner, we generated 11 keywords for each official link: the title of the corresponding Wikipedia article, plus the top 10 keywords returned by the Keyword Planner.
As examples, \Tabref{tab:keywords_example} summarizes the most relevant keywords generated for two different websites: Coursera (a popular online course platform) and American Airlines.

\begin{table}[t]
\scriptsize
\centering
\caption{Keywords, alongside estimated average cost per click (CPC), for two example websites.}
\vspace{-2mm}
\begin{tabular}{lll}
 & \thead[l]{\textbf{Coursera}} & \thead[l]{\textbf{American Airlines}} \\
 \hline
 \textbf{From title} & \cpt{{coursera}}& \cpt{{American Airlines}}\\
 \hline
\textbf{\makecell[l]{Keywords\\recommended\\by Google Ads\\Keyword Planner}} & \makecell[l]{\cpt{online courses}\\\cpt{online colleges}\\ \cpt{online classes}\\\cpt{mooc}\\\cpt{online learning}\\\cpt{free online courses}\\\cpt{online degrees}\\\cpt{open university courses}\\\cpt{online education}\\\cpt{online universities}} & \makecell[l]{\cpt{aa}\\\cpt{airline flights}\\\cpt{airline tickets}\\\cpt{airlines}\\\cpt{american}\\\cpt{american airlines flights}\\\cpt{cheap air tickets}\\\cpt{cheap airline tickets}\\\cpt{flight tickets}\\\cpt{us airways}
} \\
\hline
\textbf{Est. avg. CPC} & \$0.79 & \$1.10 \\
\hline
\end{tabular}
\label{tab:keywords_example}
\end{table}

\xhdr{Cost-per-click (CPC) forecasting}
Once the set of keywords and the bids have been set, Google Ads can make a prediction about the cost of the campaign through its forecasting tool.
The prediction model uses historical data to simulate the auction system and provides an estimated number of clicks and the average price for every keyword.
The tool predicts the campaign's average \textit{cost per click (CPC)} by combining the keyword costs with their expected \ctr{}s.
In practice, the forecasting tool simulates campaigns for specific target countries.
We used the top 5 English\hyp speaking countries (U.S., U.K., India, Canada, and Australia; \cf\ \Secref{sec:Overall click statistics}) as target countries, since they accounted for a large portion (71.6\%) of all external\hyp link clicks in the \WP logs studied here.



\xhdr{Estimating the value of official links}
Leveraging the above tools, we estimated how much a website owner would need to pay to Google Ads for a single click to their website as follows:
\begin{enumerate}
    \item Obtain keywords for the website via the Keyword Planner.
    \item For each keyword, set the bid to the 80th percentile of the keyword's historical auction price.
    \item Estimate the cost (CPC) for one click on the website link by feeding the keywords and their bids to the forecasting tool (using ``broad search'', \cf\ footnote~\ref{fn:broad search}).
\end{enumerate}

We emphasize that setting a high bid (step~2) does not automatically entail a high CPC. Indeed, as we shall see, the winning price was usually much lower than the bid. A high bid ensures that we are likely to win the simulated auction and that the promoted link is actually displayed to the user, which is required in order to obtain clicks---the event whose cost we are aiming to estimate.

\begin{figure}[t]
    \centering
    \includegraphics[width=0.95\columnwidth]{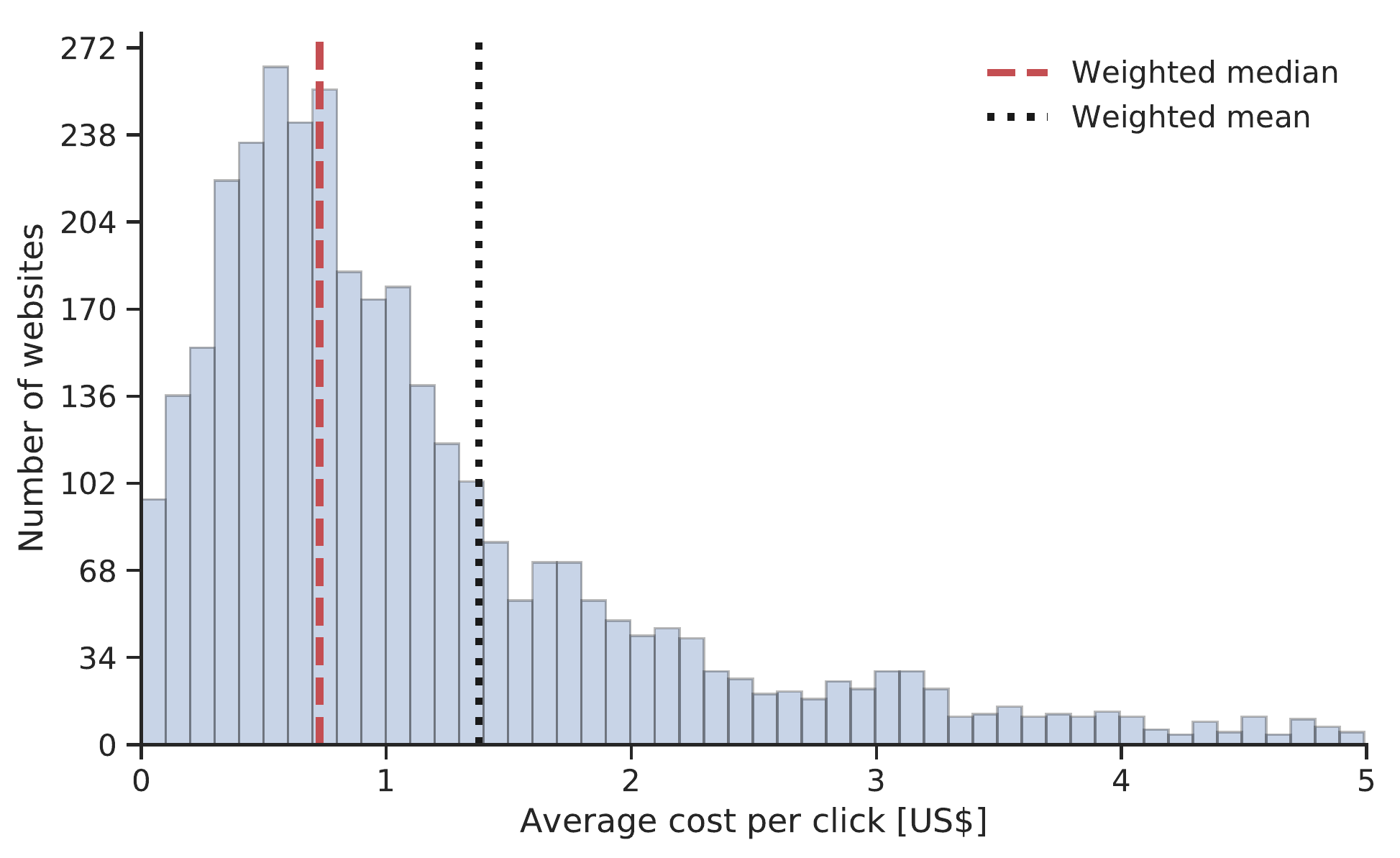}
    \vspace{-3mm}
    \caption{
Distribution of official-link CPC, estimated via Google Ads API.
Vertical lines represent weighted versions of median and mean, respectively, where each link was weighted according to its click volume in \WP logs.
    }
    \label{fig:cpc_distribution}
\end{figure}

\subsection{Results}
\label{sec:RQ3 results}

\xhdr{Cost per click (CPC)}
We applied the above-described procedure to a total of 3,600 official links from \WP infoboxes, obtained by randomly sampling an equal number of articles from each of the 57 topics.
During the one-month study period, these 3,600 official links were clicked 2.73M times in total.

As mentioned, the bid passed to the forecasting tool is not necessarily the winning price of the simulated auction; it merely caps expenses.
In practice, the auction price reached our bid in only 8.9\% of cases; on average, the auction price was 58\% of the bid.

The CPC distribution is shown in \Figref{fig:cpc_distribution}.
When weighting all links evenly (\ie, macro\hyp averaging), the mean and median CPC is \$1.64 and \$0.90, respectively.
As not all links are equally popular, a more reasonable estimate of the overall CPC may be obtained by weighting each link according to the number of clicks it received (\ie, micro\hyp averaging).
The resulting weighted CPC is slightly lower, with a mean and median of \$1.37 and \$0.73, respectively (vertical lines in \Figref{fig:cpc_distribution}).

Investigating the weighted mean CPC for individual topics, we found
considerable variation, with the highest CPCs for
\cpt{Mathematics},
\cpt{Medicine \& health},
\cpt{Books},
and \cpt{Architecture},
and the lowest CPCs for
\cpt{Music},
\cpt{Sports},
\cpt{Fashion},
and \cpt{Films}
(omitting from the list topics that mark geographical regions, such as \cpt{North America}).

\xhdr{Monthly value of traffic to official websites}
Multiplying the weighted CPC with the overall number of 9.8M clicks on official links during the one-month study period,
we estimate the total monthly value generated by the traffic from Wikipedia to official websites
as \textbf{\$13.4~million} when using the mean CPC estimate,
or as \textbf{\$7.2~million} when using the median CPC estimate.

\begin{figure*}[t]
  \includegraphics[width=0.95\textwidth]{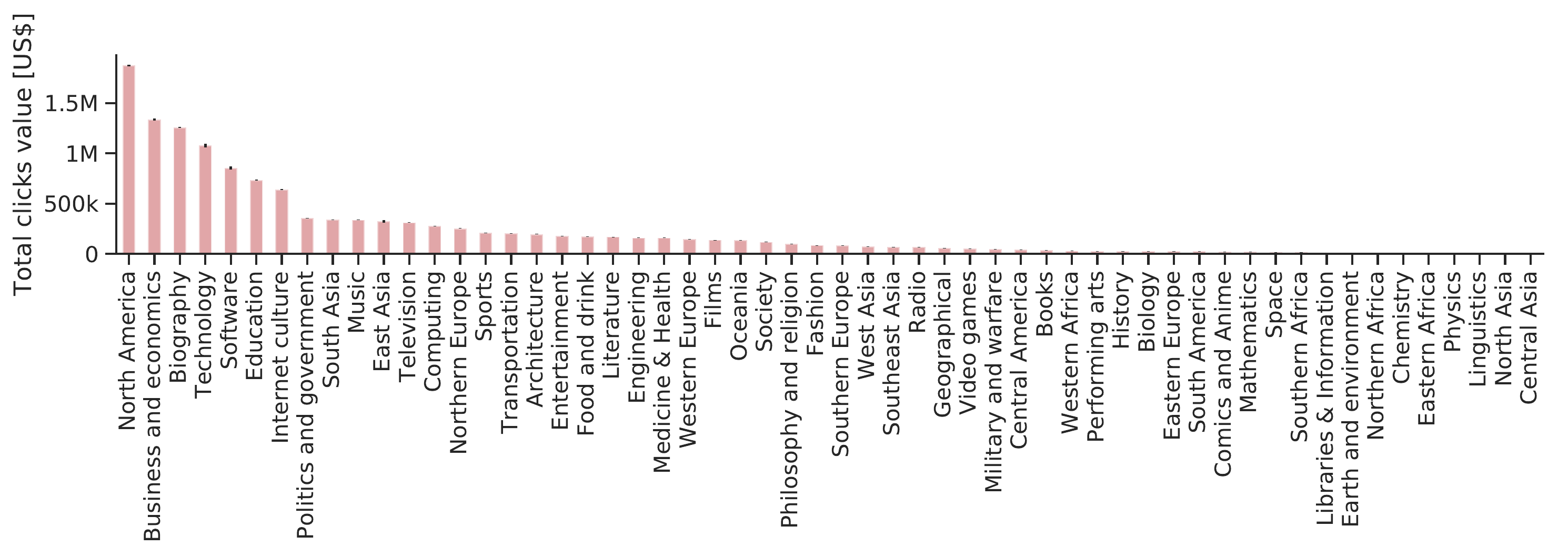}
  \vspace{-3mm}
  \caption{
  Estimated total monthly value of official links in \WP infoboxes by topic, obtained by multiplying the mean cost per click (CPC) of links from the respective topic with the total number of clicks on those links in the \WP logs.
  }
  \label{fig:total_cost_topics}
\end{figure*}

Broken down by topic,
we obtained the (mean-based) estimates of \Figref{fig:total_cost_topics}. The
topic with the highest total monthly value (\$1.9M) is \cpt{North America}, a macro category assigned to a large set of articles, including U.S.\ companies and people.
It is followed by \cpt{Business and economics} (\$1.3M),
\cpt{Biography} (\$1.3M),
\cpt{Technology} (\$1.0M),
and \cpt{Software} (\$0.9M).


\section{Discussion}
\label{sec:Discussion}

\xhdr{Wikipedia as a gateway to the Web}
While the value of Wikipedia's knowledge is fairly well known, less known was the hidden and significant additional value of Wikipedia as a gateway.
Building on top of existing work related to Wikipedia readers' behavior analysis, we have uncovered a new perspective on the role of Wikipedia in the broader Web ecosystem. 
We offered a description of the value of Wikipedia as a gateway under multiple levels of analysis. Overall, we found that a substantial fraction of Wikipedia readers use the encyclopedia as a gateway  to the broader Web: readers engage more and faster with official links in the article's infobox than with links in the article body or in the reference section. 
We found that Wikipedia's role as a gateway to external content is particularly pronounced when users visit articles about
websites, software, businesses, education, and sports, among others, where the \ctr{} of official links is the highest.

\xhdr{Wikipedia as a stepping stone}
We found an inverse relation between the time it takes to click on an external link and its average click-through rate, showing clusters of topics, such as \cpt{Sports} and \cpt{Internet culture} where engagement with links was high and fast, or conversely, where engagement with link was low and slow, such as \cpt{Biography} and \cpt{Geographical}. We also observed
that articles that were visited particularly frequently from external referrers (mostly search engines) also had a particularly high probability of an official\hyp link click after being reached from external referrers.
These results indicate that a certain distinct set of \WP articles is leveraged by users in the spirit of ``stepping stones'' or ``revolving doors'',
which are reached nearly exclusively from external referrers (mostly search) and from which the user leaves \WP immediately again by clicking an official external website link.
This begs the question: If users visit \WP from a search engine result page only to leave it immediately towards a third website, why would they not simply use the search engine to locate the third website to begin with?

We conjecture that the reason is that the search engine cannot fulfill the user's information need in such situations, whereas \WP can.
When manually screening the data
(focusing on popular articles with at least 30K pageviews during the one-month study period), we found that, among the articles with the highest CTR, there was a disproportionate fraction of file\hyp sharing (5 of the top 6) and pornographic (5 of the top 15) websites.
Such search results are frequently censored or down-ranked by search engines, depending on the search engine's corporate policy as well as legislation in the user's country.
Indeed, manually searching Google for the names of the 15 articles with the highest CTR (and more than 30K monthly pageviews) from two locations (U.S.\ and Germany)
revealed that 5 file-sharing websites and 1 pornographic website were not listed by Google among the top 10 search results in at least one of the two locations.
(Additionally, two controversial websites were not online anymore at the time of research, about 18 months after data collection.)
While these findings remain small-scale and anecdotal, they suggest that \WP fills a functional gap,
as a workaround for content suppressed by search engines (sometimes for valid reasons, \eg, when the linked material is copyrighted or illegal).

\WP's role as a transitory stepping stone towards external content can have important implications for Web user studies.
For example, researchers working on disinformation diffusion might want to take into account the function of Wikipedia as a short yet crucial stopover in Web users' information\hyp seeking journeys.

\xhdr{The economic value of traffic generated by Wikipedia}
Finally, we set out to estimate the monetary value of Wikipedia as a gateway to the broader Web.
The infoboxes contained in English \WP articles collectively list over half a million official\hyp website links, which were clicked 9.8M times during our one-month study period.
These clicks were generated by \WP for free, amidst a Web ecosystem that is majorly powered by paid ads.
We asked, ``If the respective website owners wanted to achieve the same number of clicks via sponsored search results, what would be the price?''
We estimate that achieving the 9.8M monthly clicks on official links would cost a total of \$7--13 million using Google Ads.
Extrapolating to 12 months, the yearly cost would amount to \$84--156 million.
This is a remarkably high number, considering that the annual revenue of the Wikimedia Foundation, the non-for-profit organization that operates Wikipedia and its sister projects, stands at around \$110 million,%
\footnote{\url{https://wikimediafoundation.org/about/financial-reports}}
coming entirely from donations and voluntary contributions.
We also emphasize that the estimated economic value of \$84--156 million pertains to English \WP only, whereas Wikimedia's annual revenue of \$110 million needs to support all Wikimedia projects across languages.

We showed that, when buying clicks from Google Ads instead of obtaining them from \WP for free, the types of businesses that would have to pay the most would be North American companies, as well as software and technology businesses. While the narrative about tech companies' donations to Wikipedia has often been around their massive usage of the free encyclopedic content for products and algorithms \cite{techcrunch2019,slate2018amazon},
these findings might provide yet another perspective on how these companies benefit from the hard work of hundreds of thousands of volunteer editors.

More broadly, our work expands the small body of literature on measuring the value of Wikipedia to the Web~\cite{vincent_2018,mcmahon2017substantial}. While previous work focused on the value of content \textit{production,} for example estimating that Wikipedia generates \$1.7 million of Reddit and Stack Overflow's revenue, based on the amount of Wikipedia\hyp linked posts on those platforms~\cite{vincent_2018}, we focused here on the value of Wikipedia \textit{traffic}. 
We provided, for the first time, an estimation of the monetary value offered---for free---by Wikipedia to the broader Web ecosystem by means of link navigation.


\vfill
\xhdr{Limitations and future work}
This study should be considered in the light of its limitations.
Most notably, it was constrained to data collected during one month from English \WP only, and as such provides a limited view of readers' general behavior.
Future work should replicate the study for different time periods and language versions in order to paint a more complete and inclusive picture of Wikipedia readers' engagement with external links.

Besides broadening the scope, future work should also go deeper by more closely investigating why certain types of official link see particularly high or low CTRs
(\eg, the CTR of links to geographical and biographical content was particularly low; \cf\ \Figref{fig:coefficients_ctr}).
Also, considering that official links related to \cpt{Business and economics} saw the highest CTRs, it will be interesting to analyze which businesses benefit most from the free traffic provided by \WP.

Whereas our investigation of the volume (RQ1) and patterns (RQ2) of engagement with external links on \WP was primarily measurement\hyp based, our estimation of the economic value of \WP as a gateway to the Web (RQ3) was more speculative.
On the one hand, we operated under the assumption that our methodology for obtaining costs per click via the Google Ads API is sound and provides accurate estimates, despite the fact that we relied on keyword suggestions of unknown quality from the Google Ads Keyword Planner and on uncertain auction simulations from the Google Ads forecasting tool.
On the other hand, and more fundamentally, estimating the economic value of the Web traffic generated by \WP necessarily requires arguing about a hypothetical, counterfactual (``what if'') situation, in our case,
``What if website owners were to pay for the same number of clicks via Google Ads instead of obtaining them from \WP for free?''

Although similar reasoning has been applied to estimate the value of images from Wikimedia Commons~\cite{erickson2018commons}, it remains open how realistic that ``what if'' is:
as a matter of fact, \WP \textit{is} providing those clicks for free, so why would website owners ever decide to pay for them instead?
As one concrete example, one could imagine a situation where a website owner would want to increase traffic to their site, in which case our estimates indicate how much it would cost them to double the traffic they already receive from \WP for free.
Alternatively, one could imagine scenarios where \WP were to be blocked by censorship or ceded to exist entirely, in which case our estimated economic value of traffic from \WP would correspond to the loss on behalf of website owners due to the lack of that traffic.
Finally, and more boldly, one could imagine a setting where \WP decided to introduce a fee for clicks on official website links, in which case our estimate would upper-bound the amount of extra revenue \WP could possibly earn from such a fee.
Although the latter setting is highly unrealistic, we consider it a useful thought experiment that can help emphasize Wikipedia's importance as a provider of free traffic.

As a final remark, we argue that all notions of economic value are fundamentally counterfactual at heart, as they always consider ``what if'' scenarios (``If A were to give X to B, how much money would B give to A in return?''),
which is also the reason why business valuations of companies are routinely criticized as absurdly off~\cite{ft2020}.

To conclude, we hope this work will offer ideas and methods to those interested in exploring \WP's role in the larger Web ecosystem in more depth, and that it will help quantify the true value of the largest encyclopedic knowledge repository on the Web.


{\small
\xhdr{Acknowledgments}
West's lab was partly funded by
Swiss National Science Foundation (grant 200021\_185043),
European Union (TAILOR, grant 952215),
Microsoft Swiss Joint Research Center,
and by generous gifts from Facebook and Google.
We would also like to thank Leila Zia and Joseph Seddon from Wikimedia for providing valuable suggestions for this research. 
}

\balance

\bibliographystyle{ACM-Reference-Format}
\bibliography{main}

\end{document}
\endinput